\documentclass[11pt]{article}

\usepackage[preprint]{acl}

\usepackage{times}
\usepackage{latexsym}
\usepackage{subfigure}
\usepackage{subcaption}
\usepackage{times}  % DO NOT CHANGE THIS
\usepackage{helvet}  % DO NOT CHANGE THIS
\usepackage{courier}  % DO NOT CHANGE THIS
\usepackage{algorithm}
\usepackage{algorithmic}
\usepackage[most]{tcolorbox}
\usepackage{booktabs}
\usepackage[table]{xcolor}
\usepackage{caption}
\usepackage{xcolor}
\usepackage{tabularx}

\usepackage{graphicx}
\usepackage{natbib}  % DO NOT CHANGE THIS AND DO NOT ADD ANY OPTIONS TO IT
\usepackage{tabularx}
\usepackage{caption} % DO NOT CHANGE THIS AND DO NOT ADD ANY OPTIONS TO IT
\usepackage{algorithm}
\usepackage{algorithmic}
\usepackage{amsmath}
\usepackage{amssymb}
\usepackage{multirow}
\usepackage{booktabs}
\renewcommand{\arraystretch}{1.4}
\usepackage[table]{xcolor}
\usepackage{caption}
\usepackage{xcolor}

\usepackage[T1]{fontenc}
\usepackage[utf8]{inputenc}

\usepackage{microtype}

\usepackage{inconsolata}

\usepackage{graphicx}

\title{A Progressive Evaluation Framework for Multicultural Analysis of Story Visualization}

\author{
  Janak Kapuriya$^{\ast}$ \hspace{1em}  Ali Hatami \hspace{1em} Paul Buitelaar \\
  Data Science Institute, University of Galway, Galway, Ireland \\
  \{janakkumar.kapuriya, ali.hatami, paul.buitelaar\}@insight-centre.org \\ 
  $^\ast$Corresponding Author
}

\begin{document}

\maketitle

\begin{abstract} 

Recent advancements in text-to-image generative models have improved narrative consistency in story visualization. However, current story visualization models often overlook cultural dimensions, resulting in visuals that lack authenticity and cultural fidelity. In this study, we conduct a comprehensive multicultural analysis of story visualization using current text-to-image models across multilingual settings on two datasets: FlintstonesSV and VIST. To assess cultural dimensions rigorously, we propose a Progressive Multicultural Evaluation Framework and introduce five story visualization metrics, Cultural Appropriateness, Visual Aesthetics, Cohesion, Semantic Consistency, and Object Presence, that are not addressed by existing metrics. We further automate assessment through an MLLM-as-Jury framework that approximates human judgment. Human evaluations show that models generate more coherent, visually appealing, and culturally appropriate stories for real-world datasets than for animated ones. The generated stories exhibit a stronger alignment with English-speaking cultures across all metrics except Cohesion, where Chinese performs better. In contrast, Hindi ranks lowest on all metrics except Visual Aesthetics, reflecting real-world cultural biases embedded in current models. This multicultural analysis provides a foundation for future research aimed at generating culturally appropriate and inclusive visual stories across diverse linguistic and cultural settings.

\end{abstract}

\begin{figure}[ht]
  \centering
  \includegraphics[width=0.48\textwidth]{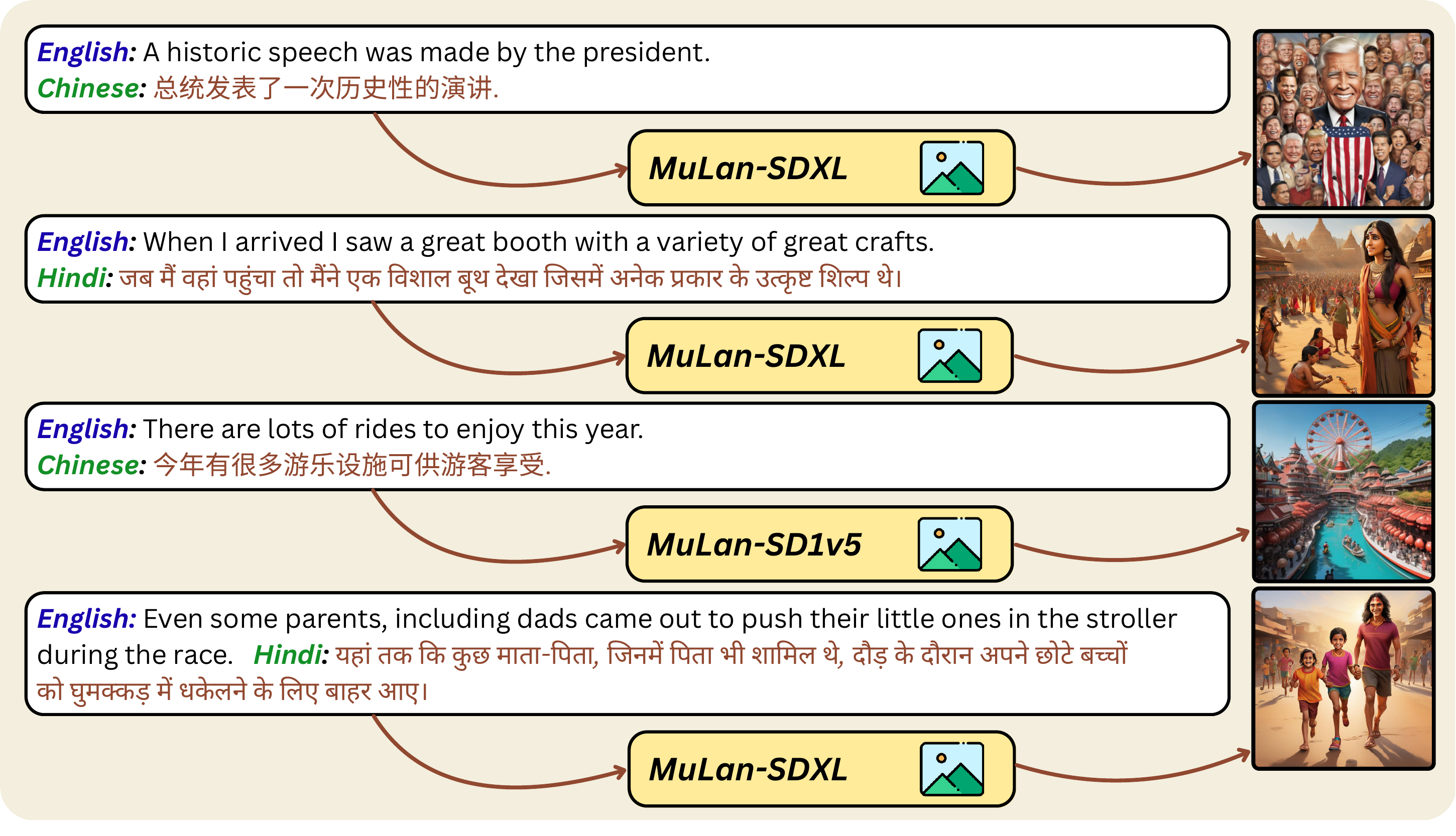}
\caption{Cultural inconsistencies and stereotypes in generated story scenes across models and languages. \textbf{Ex–1}: Model is not interpreting the word 'president' in Chinese as 'party leader' in China, and generated an image of a US president. \textbf{Ex–2}: Instead of a modern craft fair, it depicts temples and a crowded assembly of saints, reinforcing the stereotype of an ancient \textit{Mahakumbh} style fair. \textbf{Ex–3}: Red lantern stereotypes dominate the rides, misrepresenting Chinese culture. \textbf{Ex–4}: A Hindi cultural stereotype depicts a parent leading children in a race rather than using a stroller, with several children barefoot, reinforcing inaccurate cultural assumptions.}
  \label{fig:cultural_issues}
\end{figure}

\section{Introduction}
Recent advancements in text-to-image diffusion models, such as DALLE-3 \citep{betker2023improving}, SDXL \citep{podell2023sdxl} and Imagen 3 \citep{baldridge2024imagen} have achieved remarkable performance in generating photorealistic images from text prompts. These models have set new benchmarks for producing high-quality images, opening up a wide array of real-world applications including controlled image editing ControlNet \citep{zhang2023adding}, FLUX.1 \citep{labs2025flux1kontextflowmatching} and video generation from textual prompts SORA \citep{videoworldsimulators2024}, Stable Video Diffusion \citep{blattmann2023stable} and Veo 2 \citep{veo2024}. Within this broader landscape, one promising and rapidly growing application is Story Visualization, which focuses on generating coherent sequences of visual scenes based on sequences of storylines. This task holds immense potential across fields such as education, animation and content creation. However, current story visualization models are predominantly advanced in English, with their capabilities in multilingual settings remaining largely unexplored.

% \begin{figure}[ht]
%   \centering
%   \includegraphics[width=0.5\textwidth]{cultural_issue.pdf}
%   \caption{This figure highlights cultural inconsistencies in image generation across different models. \textbf{Ex-1} shows faces with English cultural features despite the input text being in Chinese. \textbf{Ex-2} illustrates the generation of Chinese stereotypes, such as seafood bones for a dog. \textbf{Ex-3} presents a mismatch where Hindi input results in faces resembling South African culture. \textbf{Ex-4} depicts Hindi input with the name 'Wilma' results in a generated image of a dog instead of a human.}
%   \label{fig:CapSFT}
% \end{figure}

Stories are often connected with the culture and geography they represent, serving as reflections of the traditions, values and identities of specific communities \citep{bruner2010narrative}. Despite this, current story visualization models \citep{zhou2024storydiffusion, mao2024story, yang2024seed, zheng2024temporalstory} generate sequences of images without adequately considering cultural dimensions, produces visuals that lack authenticity and cultural fidelity. This is critical in multilingual settings, where distinct linguistic backgrounds carry unique cultural nuances as shown in Figure \ref{fig:cultural_issues}. Addressing these aspects remains underexplored, leaving a gap in meaningful, accurate multilingual story visualization.

% As a result, it often overlook deeper cultural context, producing visual sequences that lack the authenticity needed to convey the traditions, values and identities at the heart of the stories they depict. The importance of cultural fidelity becomes even more pronounced in multilingual settings, where stories from diverse linguistic backgrounds inherently encompass distinct cultural nuances. Addressing these cultural aspects is crucial for creating meaningful and accurate visualizations that resonate with audiences across different regions. However, the integration of cultural elements in story visualization remains underexplored, leaving a significant gap in ensuring cultural aspects in multilingual story visualization.

Building on this, we systematically conducted a multicultural analysis of story visualization across three languages English, Chinese and Hindi on the animated story dataset FlintstonesSV \citep{gupta2018imaginethisscriptscompositions} and the real world story dataset VIST \citep{huang2016visual}, with five multilingual text-to-image generation models. To enable an in-depth evaluation of story visualization we propose five story visualization evaluation metrics such as Cohesion, Cultural Appropriateness, Object Presence, Visual Aesthetics and Semantic Consistency, which are not captured by existing evaluation methods like FID \cite{heusel2017gans} and CLIP \citep{radford2021learning}. FID and CLIP metrics focus only on image-to-image or image-to-text, whereas our proposed metrics are designed for the evaluation of story visualization in regard to text-to-image sequences and cultural aspects. To more rigorously assess cultural aspects, we introduce the Progressive Multiculture Evaluation Framework, which provides a deeper evaluation of cultural fidelity in generated stories. Since human evaluation of proposed metrics at large scale is time-consuming and costly, we utilize the MLLM-as-Jury Evaluation Framework. This framework uses three Multimodal Large Language Model (MLLM) as judges. The final rating for a story sample is determined by aggregating the ratings from the three judge models.

To perform the multicultural analysis of story visualization in multilingual settings, we formulated the following research questions:

\textbf{RQ 1:} How does the visualization of the same story varies across different languages?

\textbf{RQ 2:} How does story visualization adapt to different cultures?
\\
\textbf{Our contributions are as follows:}
\begin{itemize}
    \item \textbf{Exploration of Cultural Aspects:} We conducted a comprehensive study of the cultural dimensions of story visualization in a multilingual setting.
    \item \textbf{Proposed Story Visualization Metrics:} We propose five evaluation metrics: \texttt{Cohesion, Visual Aesthetics, Cultural Appropriateness, Object Presence, Semantic Consistency} tailored to capture the unique characteristics of story visualization, addressing gaps left by traditional evaluation metrics like FID  and CLIP.
    \item \textbf{Progressive Multicultural Evaluation Framework:} This framework enables in-depth evaluation of culture by incorporating culture relevant focus-points and illustrative examples during the evaluation process.
    \item \textbf{MLLM-as-Jury Evaluation Framework:} We introduce the MLLM-as-Jury Evaluation Framework to enable large-scale, automated evaluation of story aspects using Expert Role-specific evaluation prompts.
    \item \textbf{Multicultural Analysis on Animated and Real Stories:} We analyzed story visualization on open-ended real-world story and animated story datasets.
\end{itemize}

\section{Related Work}

\subsection{Story Visualization}
Recent advancements in story visualization have improved narrative consistency through several dataset-specific training approaches. These include models such as Make-A-Story \citep{rahman2023make}, StoryGPT-V \citep{shen2023large}, ARLDM \citep{Pan_2024_WACV}, and TemporalStory \citep{zheng2024temporalstory}. Other work explores long‑story visualization \citep{yang2024seed} and training‑free approaches like StoryDiffusion \citep{zhou2024storydiffusion} and StoryAdapter \citep{mao2024story}. However, these methods focus on English scene descriptions and overlook the cultural dimension in story visualization. To address this gap, we conducted a multicultural analysis of story visualization in multilingual contexts using multilingual text-to-image models.

\subsection{Multicultural Analysis of VLMs and Text-to-Image Models}
Recent research has increasingly examined cultural aspects across large language models (LLMs), vision–language models (VLMs), and text-to-image (T2I) systems. \citet{adilazuarda2023culture} highlights gaps in cultural modelling within LLMs, while \citet{liu2023culturalgap} and \citet{ventura2023culturalpov} demonstrate cultural inconsistencies in T2I generation and propose methods to adjust outputs according to cultural perspectives. \citet{khanuja-etal-2024-image} propose an approach to directly align generated images with target cultural contexts. In the visual question answering domain, the \textsc{CVQA} and \textsc{cultureVQA} datasets show that VLMs perform poorly on culture-centric questions \citep{romero2024cvqa, nayak2024benchmarking}, and other studies further assess cultural sensitivity in VLMs \citep{bhatia2024multicultural, yadav2025cultural, liu2025culturevlm}. \citet{bayramli2025cultdiff} report that generated images can reinforce stereotypes or misrepresent cultural elements, while \citet{bhalerao2025mosaig} proposes a multi-agent framework to produce culturally relevant imagery. In contrast to these works, our work focuses on a detailed multicultural analysis of story visualization using text-to-image models, aiming to uncover how cultural nuances specifically influence narrative visual generation across languages.

\begin{figure*}[ht]
  \centering
  \includegraphics[width=\textwidth]{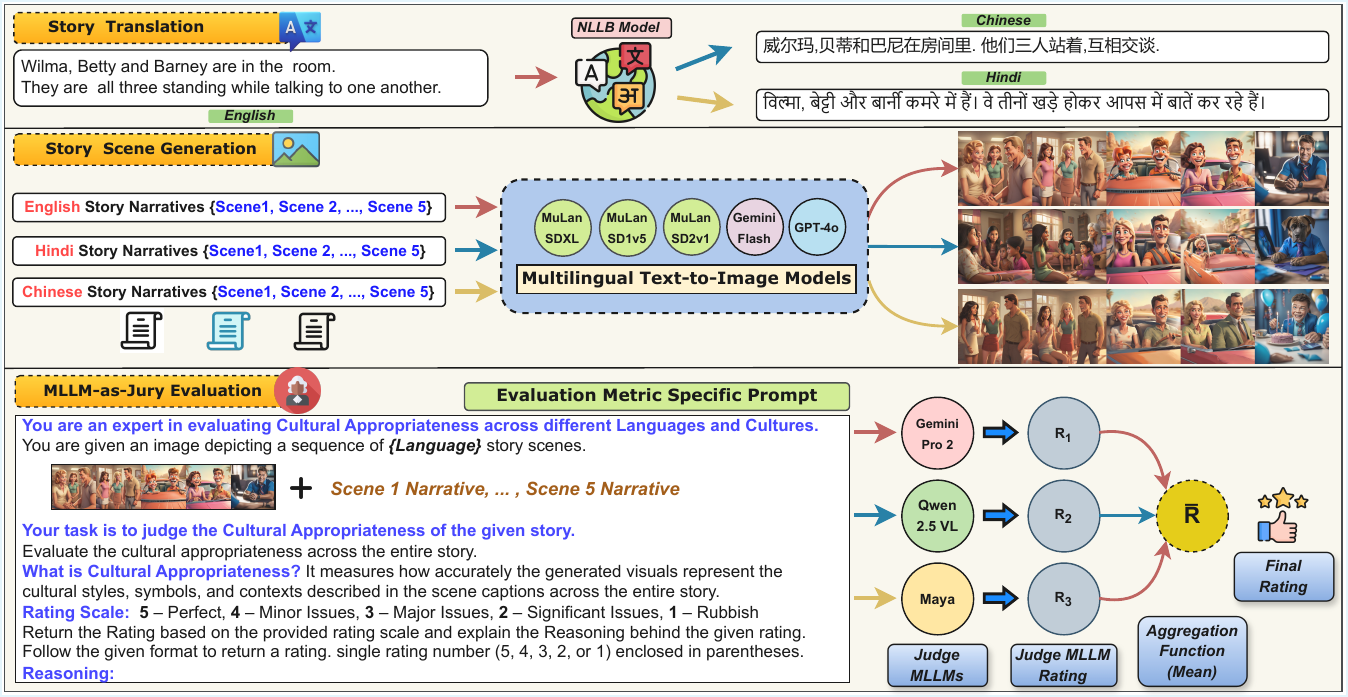}
  \caption{Three-stage framework for multicultural analysis of story visualization using text-to-image models.}
  \label{fig:multucultural_analysis_workflow}
\end{figure*}

\subsection{MLLM-as-Judge Models}
Recent work in automatic evaluation has adopted the LLM‑as‑Judge paradigm, using LLMs to assess generated outputs in domain‑specific tasks \citep{zhu2023judgelm, kocmi-federmann-2023-large, chiang2023can} while \citep{verga2024replacing} proposed LLM-as-Jury by aggregating judge scores from distinct families of judge models to reduce bias of single judge LLM. \citep{chen2024mllm} extended the LLM-as-Judge approaches to the multimodal case, applying pairwise and batch‑wise ranking across diverse datasets. However, single MLLMs-as-judge still exhibit biases in their scores. Furthermore, \citep{lee2024prometheus} observed that MLLMs acting as judges occasionally assign disproportionately high scores, exacerbating the bias. To overcome these issues, we extend LLM-as-Judge to LLM-as-Jury in our evaluation of multicultural story visualization. we adopt an MLLM‑as‑Jury framework that aggregates judgments from three different families of MLLM judges, reducing individual bias and enhancing robustness of the evaluation process.

% Recent advancements in the automatic evaluation of domain-specific problems leverage the LLM-as-Judge framework \citep{zhu2023judgelm, kocmi-federmann-2023-large, chiang2023can}, where LLMs are employed to evaluate generated responses. Similarly, in the multimodal domain, \citep{chen2024mllm} proposed the MLLM-as-Judge framework to assess the capabilities of various multimodal large language models (MLLMs) across diverse judgment task like pairwise and batch-wise ranking across multiple datasets. However, their findings highlight persistent biases in the judgments of single-judge MLLM models. Furthermore, \citep{lee2024prometheus} observed that MLLM acting as judges occasionally assign disproportionately high scores, exacerbating the bias. To address these issues in the evaluation of multicultural aspects of story visualization  we adopt the MLLM-as-Jury evaluation framework, which aggregates judgments from three different MLLM judge models using an aggregation function. This approach mitigates bias from single-judge models and enhances the robustness of the evaluation process.

\section{Methodology}

\subsection{Story Translation}

To support multilingual story visualization, each story narrative \( s_i^{(\text{English})} \), originally written in English, is translated to Hindi and Chinese as shown in stage 1 of Figure \ref{fig:multucultural_analysis_workflow}, using the \texttt{NLLB-200} \cite{nllbteam2022languageleftbehindscaling} translation model. This results in a multilingual set of narratives \( s_i^{(\ell)} \) for each language \( \ell \in \mathcal{L} = \{ \text{English}, \text{Hindi}, \text{Chinese} \} \). These translated narratives serve as inputs for the subsequent multilingual T2I generation. A detailed analysis of the translation quality, including reference-free COMET-QE \cite{rei-etal-2020-comet} scores for both datasets, is provided in Appendix~\ref{ref_free_translation}.
% , enabling cross-lingual consistency and cultural analysis in visual storytelling.

\subsection{Story Visualization using Multilingual T2I Models}
%Each story narrative \( s_i^{(\text{English})} \), originally written in English, is translated into additional languages \( \ell \in \{ \text{Hindi}, \text{Chinese} \} \), resulting in multilingual versions \( s_i^{(\ell)} \) for each \( \ell \in \mathcal{L} = \{ \text{English}, \text{Hindi}, \text{Chinese} \} \). 

% Let \( M \) denote a multilingual text-to-image (T2I) model. The image corresponding to each narrative is generated with the function: \[
% I_i^{(\ell)} = \mathcal{F}(M, s_i^{(\ell)}, T, \gamma), \quad \forall \ell \in \mathcal{L},
% \]
% where \( I_i^{(\ell)} \) is the image generated from the narrative \( s_i^{(\ell)} \) in language \( \ell \), using model \( M \) with \( T \) representing the number of generation steps in diffusion denoising steps and \( \gamma \) denoting the guidance scale.

\noindent As shown in stage 2 of Figure \ref{fig:multucultural_analysis_workflow}, Let $M$ denote a multilingual T2I model. We generate the image corresponding to each narrative $s_i^{(\ell)}$ in language $\ell$ independently\footnote{We generated each scene independently due to unavailability of multilingual story visualization models and our primary objective is to analyze culture rather than cross-scene consistency.} using the function $I_i^{(\ell)} = \mathcal{F}(M, s_i^{(\ell)}, T, \gamma)$ for all $\ell \in \mathcal{L}$, where $T$ is the number of denoising steps and $\gamma$ is the guidance scale.

\subsection{Proposed Story Visualization Evaluation Metrics} 
To evaluate the quality of the generated text-to-image sequences, which are not captured by existing metrics FID and CLIP, we propose two types of evaluation metrics: \textbf{(1) Sequential Scene Evaluation} and \textbf{(2) Individual Scene Evaluation}. 

%Sequential evaluation metrics assess the story as a whole by considering all scene images and their corresponding narratives. In contrast, individual scene evaluation metrics focus on evaluating each image-narrative pair independently. In Sequential Scene Evaluation, the evaluation is based on the entire image sequence, with a single score from 1 to 5 assigned to reflect the overall quality. In contrast, Individual Scene Evaluation involves assessing each image separately, resulting in five individual scores per image. The following guideline is used for scoring: 5 – Perfect, 4 – Minor Issues, 3 – Major Issues, 2 – Significant Issues and 1 – Rubbish.

% The metrics used for each category are as follows:
\begin{itemize}
    \item \textbf{Sequential Scene Evaluation:}
    \begin{itemize}
        \item \textbf{Cultural Appropriateness}: Measures how accurately the generated visuals represent the cultural styles, symbols and contexts described in the scene captions over the full story.
        \item \textbf{Cohesion}: Measures the coherence of visual elements across all scenes in a story, ensuring smooth transitions and logical relationships between scenes.
        \item \textbf{Visual Aesthetics}: Evaluates the overall appeal of the visuals in terms of clarity, color, and sharpness over the full story.
    \end{itemize}
    \item \textbf{Individual Scene Evaluation:}
    \begin{itemize}
        \item \textbf{Object Presence}: Evaluates the inclusion of all essential objects mentioned in the captions for each scene.
        \item \textbf{Semantic Consistency}: Assesses how well the visual content aligns with the semantic meaning and details provided in the text descriptions for each scene.
    \end{itemize}
\end{itemize}

Sequential evaluation metrics assess the entire story by considering all image-narrative pairs and assigning a single score (1–5) for overall quality. In contrast, individual scene metrics evaluate each pair independently, yielding five separate scores per story. We then take an average of five scores as the final score. \textbf{Rating Scheme}: \textbf{5} – Perfect, \textbf{4} – Minor Issues, \textbf{3} – Major Issues, \textbf{2} – Significant Issues, \textbf{1} – Rubbish.

%Sequential evaluation metrics assess the story as a whole by considering all scene images and their corresponding narratives, assigning a single score (1 to 5) to reflect overall quality. In contrast, individual scene evaluation metrics assess each image-narrative pair independently, resulting in five separate scores per story. The scoring guideline is as follows: 5 – Perfect, 4 – Minor Issues, 3 – Major Issues, 2 – Significant Issues and 1 – Rubbish.

\subsection{Progressive Multiculture Evaluation Framework}

To perform a detailed evaluation of cultural aspects in generated story images, this framework incrementally enriches the cultural evaluation criteria across three levels:

\begin{itemize}
    \item \textbf{Version 1 (V1)}: Culture Definition $(\mathcal{C})$ - evaluates culture appropriateness based on the just definition.
    \item \textbf{Version 2 (V2)}: Culture Definition $(\mathcal{C})$ + Focus Points $(\mathcal{F})$ - introduces specific cultural dimensions to pay attention to during the evaluation.\\
    \textbf{Focus Points $(\mathcal{F})$:} Background Objects, Facial Features, Infrastructures, Apparel
    \item \textbf{Version 3 (V3)}: Culture Definition $(\mathcal{C})$ + Focus Points $(\mathcal{F})$ + Illustrative Examples $(\mathcal{E})$ - provides concrete examples for each focus point to guide consistent and precise assessment.\\
    \textbf{Illustrative Examples $(\mathcal{E})$}\\
    \textbf{1. Background Objects}: Assess whether the depicted objects represent the target culture setting described in the scenes, focusing on nearby objects, furniture, decorations and other contextual details.\\
\textbf{2. Character's Facial Features}: Evaluate whether the facial structures align
with the diverse traits commonly found in the target culture. Avoid assumptions
about stereotypical features.\\
\textbf{3. Infrastructures}: Consider whether the settings, such as architectural
elements, are appropriate for the target culture. \\
\textbf{4. Apparel}: Assess whether the clothing aligns with traditional or contemporary styles representative of the Appropriate culture.

\end{itemize}

Formally, let $S = \{s_1, s_2, \ldots, s_n\}$ denote the set of $n$ generated story samples. Each story $s_i$ is evaluated using a progressively detailed evaluation mapping $\mathbb{E}_v(s_i)$, where $v \in \{1, 2, 3\}$ represents the evaluation version:
\vspace{-0.6em}  % adjust value as needed
\[
\begin{aligned}
\mathbb{E}_1(s_i) &= g(s_i \mid \mathcal{C}) \\
\mathbb{E}_2(s_i) &= g(s_i \mid \mathcal{C}, \mathcal{F}) \\
\mathbb{E}_3(s_i) &= g(s_i \mid \mathcal{C}, \mathcal{F}, \mathcal{E})
\end{aligned}
\]
\vspace{-1em}

The goal of this progressive structure is to evaluate whether richer prompts lead to more accurate and rigorous culture assessments of the generated story content.

% This progressive structure allows us to analyze how added context (focus points and examples) affects the consistency and granularity of cultural evaluation. The goal is to see whether richer prompts lead to more accurate and culturally sensitive assessments of generated story content.

%This progressive structure allows us to analyze how additional context (focus points and examples) influences the consistency and granularity of cultural evaluation. Our goal is to determine whether richer prompts lead to more accurate and culturally sensitive assessments of generated story content.

\subsection{MLLM-as-Jury Evaluation Framework}

% To enable large‑scale, automatic assessment of different story aspects, we propose a \emph{MLLM‑as‑Jury} evaluation framework that leverages Role‑specific Chain‑of‑Thought (CoT) prompting. Each story is represented as a sequence of narrative and scene image pairs,
% \[
% S = \{(n_1, i_1), (n_2, i_2), \dots, (n_T, i_T)\},
% \]
% where \(n_t\) is the narrative and \(i_t\) is the generated image from multilingual T2I models for \(t^\text{th}\) scene. A Role‑specific CoT prompt is constructed from function \(\mathcal{P}\) from an Role-specific evaluation metric instruction and story given by $p = \mathcal{P}(\text{Instruct}, S) $
% and then fed to three different MLLM judge models:
% \[
% \mathcal{M}_1 = \mathrm{J}_{\mathrm{Gemini}}, \quad
% \mathcal{M}_2 = \mathrm{J}_{\mathrm{Qwen2.}}, \quad
% \mathcal{M}_3 = \mathrm{J}_{\mathrm{Maya}}.
% \]
% Each judge produces a rating for the evaluation metric, denoted \(r_1 = \mathcal{M}_1(p)\), \(r_2 = \mathcal{M}_2(p)\) and \(r_3 = \mathcal{M}_3(p)\).
% The final rating is computed by mean aggregation function over three judges rating given by  \[
% R_{pred} = \frac{1}{3}\bigl(r_1 + r_2 + r_3\bigr).
% \]  
% This ensemble aggregation mitigates individual model biases and enhances the robustness of the evaluation process.

To enable large-scale, automatic assessment of story visualization metrics, we propose a \emph{MLLM-as-Jury} evaluation framework that leverages Expert Role-specific prompting\footnote{MLLM-as-Jury evaluation prompts are given in the Appendix.} shown in stage 3 of Figure \ref{fig:multucultural_analysis_workflow}. Each story is represented as a sequence of narrative and scene image pairs,  $ S = \{(n_1, i_1), (n_2, i_2), \dots, (n_T, i_T)\}, $ where $n_t$ is the narrative and $i_t$ is the generated image from multilingual T2I models for the $t^\text{th}$ scene. An Expert Role-specific prompt $\mathcal{P}$ fed to three different MLLM judge models:  
$$
\mathcal{M}_1 = J_{\mathrm{Gemini}}, \quad
\mathcal{M}_2 = J_{\mathrm{Qwen}}, \quad
\mathcal{M}_3 = J_{\mathrm{Maya}}.
$$
Each judge model produces a rating for the particular evaluation metric, denoted $r_1 = \mathcal{M}_1(P)$, $r_2 = \mathcal{M}_2(P)$ and $r_3 = \mathcal{M}_3(P)$. The final rating is computed by a mean aggregation function over the three judges' ratings given by  $
R_{\mathrm{pred}} = \tfrac{1}{3}\bigl(r_1 + r_2 + r_3\bigr).$ 

We used the mean as an aggregation function rather than the median because the median removes extreme scores, which can unintentionally over-weight a single model’s behavior. We focus here on integrating different judge models' complementary perspectives to mitigate overall bias in assessment.

The three models were selected for their complementary multimodal and multilingual strengths. \textbf{Gemini-Pro 2.0} is strong in multimodal multilingual reasoning, while \textbf{Qwen2.5-VL} \citep{bai2025qwen2}
excels at image understanding and multilingual instruction-following, and \textbf{Maya} \citep{alam2024maya} performs well in multilingual and culturally sensitive tasks. 

% To evaluate generated stories, we use the \textit{VLM-as-Jury} framework, which allows automated assessment with Multimodal Large Language Models (MLLMs). We select three diverse MLLMs: \textbf{1)} \texttt{Gemini Pro 2.0}, \textbf{2)} \texttt{Qwen2.5-VL} \citep{bai2025qwen2} and \textbf{3)} \texttt{Maya} \citep{alam2024maya}.  Gemini Pro 2.0 is a SOTA model with strong multimodal multilingual capabilities, Qwen2.5-VL excels at image understanding and multilingual instruction-following and Maya performs well in multilingual and culturally sensitive tasks. Together, their strengths support robust evaluation in our MLLM-as-Jury evaluation framework.

\subsection{Human Evaluation Process}
To assess the quality and cultural relevance of generated story images, we conducted a human evaluation with native English, Chinese, and Hindi speakers. Three fluent native speakers per language (a total of nine) volunteered from the research community. Their linguistic and cultural backgrounds ensured reliable evaluation. Five male and four female evaluators participated, offering diverse perspectives. Each received 25 randomly selected story samples, consisting of five sequences of (image, caption) pairs spanning different models and datasets. Evaluation followed the Sequential and Individual Scene criteria. In addition to ratings, evaluators provided qualitative feedback, supporting a nuanced analysis of visual and cultural effectiveness in multilingual story generation. Further details on agreement scores and evaluation guidelines are provided in Appendix~\ref{human_eval_appendix}.

\renewcommand{\arraystretch}{1.1}
%\Large{
\begin{table*}[ht]
\centering
\resizebox{\textwidth}{!}{%
\begin{tabular}{l
  *{3}{c}
  *{3}{c}
  *{3}{c}
  *{3}{c}
  *{3}{c}}
\toprule
\multicolumn{1}{l}{\Large\textbf{Models}}
  & \multicolumn{3}{c}{\Large\textbf{Cultural Appr. (V3) $\uparrow$}}
  & \multicolumn{3}{c}{\Large\textbf{Visual Aesthetics $\uparrow$}}
  & \multicolumn{3}{c}{\Large\textbf{Cohesion $\uparrow$}}
  & \multicolumn{3}{c}{\Large\textbf{Semantic Consistency $\uparrow$}}
  & \multicolumn{3}{c}{\Large\textbf{Object Presence $\uparrow$}} \\
\cmidrule(lr){2-4} \cmidrule(lr){5-7} \cmidrule(lr){8-10} \cmidrule(lr){11-13} \cmidrule(lr){14-16}
\multicolumn{1}{l}{}
  & \large\textbf{English} & \large\textbf{Hindi} & \large\textbf{Chinese}
  & \large\textbf{English} & \large\textbf{Hindi} & \large\textbf{Chinese}
  & \large\textbf{English} & \large\textbf{Hindi} & \large\textbf{Chinese}
  & \large\textbf{English} & \large\textbf{Hindi} & \large\textbf{Chinese}
  & \large\textbf{English} & \large\textbf{Hindi} & \large\textbf{Chinese} \\
\rowcolor{gray!25}
\multicolumn{16}{c}{\large\textbf{FlintstonesSV}} \\
\large MuLan SD2v1     
  & 3.84 & 3.18 & 3.09
  & 3.79 & 3.73 & 3.76
  & 3.31 & 3.11 & 3.33
  & 4.02 & 3.46 & 3.89
  & 4.46 & 4.05 & 4.47 \\
\large MuLan SD1v5     
  & 3.86 & 3.21 & 3.16
  & 3.86 & 3.76 & 3.84
  & 3.35 & 3.19 & 3.35
  & 4.09 & 3.58 & 3.83
  & 4.46 & 4.15 & 4.44 \\
\rowcolor{orange!15}
\large MuLan SDXL     
  & \textbf{3.87} & \textbf{3.40} & \textbf{3.18}
  & \textbf{4.05} & \textbf{3.97} & \textbf{4.01}
  & \textbf{3.35} & \textbf{3.19} & \textbf{3.43}
  & \textbf{4.15} & \textbf{3.72} & \textbf{4.05}
  & \textbf{4.50} & \textbf{4.20} & \textbf{4.52} \\
\addlinespace
\rowcolor{gray!25}
\multicolumn{16}{c}{\large\textbf{VIST}} \\
\large MuLan SD2v1     
  & 3.94 & 3.45 & 3.49
  & 3.96 & 3.84 & 3.94
  & 3.70 & 3.39 & 3.46
  & 4.29 & 3.83 & 3.76
  & 4.57 & 4.46 & 4.45 \\
\large MuLan SD1v5     
  & 4.03 & 3.82 & 3.90
  & 4.02 & 3.89 & 4.06
  & 3.81 & 3.51 & 3.59
  & 4.37 & 4.02 & 3.96
  & 4.59 & 4.45 & 4.49 \\
  \rowcolor{orange!15}
\large MuLan SDXL      
  & \textbf{4.12} & \textbf{3.89} & \textbf{3.95}
  & \textbf{4.16} & \textbf{4.03} & \textbf{4.13}
  & \textbf{3.90} & \textbf{3.58} & \textbf{3.69}
  & \textbf{4.46} & \textbf{4.18} & \textbf{4.16}
  & \textbf{4.65} & \textbf{4.56} & \textbf{4.59} \\
\bottomrule
\end{tabular}%
}
\caption{Results of MLLM-as-Jury across three different models on 500 test samples on two datasets across multiple evaluation metrics and languages with score range (1-5).}
\label{tab:MLLM_as_Jury_results}
\end{table*}
%}

% ------------------

%}

\section{Experiments}

\subsection{Choices of Language and Datasets}
To analyze the cultural aspects of story visualization, we selected three languages: English (original language of the datasets used), Hindi and Chinese. These choices were guided by several factors, like representation of distinct geographical and cultural regions, the availability of high-quality translation models, compatibility of MLLM used in the MLLM-as-Jury framework and accessibility of annotators for human evaluation.

We analyze story visualization across cultural contexts with two datasets: VIST (real-world) and FlintstonesSV (animated). We performed experiments on randomly sampled 500 stories from each dataset, where each story has 5 (image, caption) pairs. 
VIST \citep{huang2016visual} features photo sequences from Flickr albums of everyday events, consisting of 50,000 stories and FlintstonesSV \citep{gupta2018imaginethisscriptscompositions}, based on the classic animated series "\textit{The Flintstones}", includes over 24,000 image-caption pairs focused on seven main characters in varied scenes.

\subsection{Multilingual Text-to-Image Models}
To generate story images from scene narratives in different languages, we use the multilingual T2I diffusion model MuLan\footnote{Model Configuration for all experiments is provided in the Appendix} \citep{xing2024MuLan}. We use three MuLan variants: \textbf{1)} \texttt{SD1v5}, \textbf{2)} \texttt{SD2v1} and \textbf{3)} \texttt{SDXL}, representing progressively diverse open-source stable diffusion versions along with two closed-source models, \textbf{4)} \texttt{GPT-4o} and \textbf{5)} \texttt{Gemini-Flash 2.5}, known for their strong image generation capability across languages.

\section{Results}

% This section presents a analysis of different story aspects on our propose five evaluation metrics, using five models evaluated on two datasets across three languages.\\

\noindent \textbf{MLLM-as-Jury results on Test set:} As shown in Table~\ref{tab:MLLM_as_Jury_results}, SDXL consistently outperforms SD2v1 and SD1v5 across all metrics, languages, and datasets. Scores are generally higher on VIST than on FlintstonesSV, for all three languages, indicating that models handle real-world content better than animated scenes. English receives the highest scores overall, while Chinese and Hindi lag behind, with Hindi performing the lowest in Cohesion and Cultural Appropriateness.

Table~\ref{tab:Average_Results_Across_Metrics} shows that, when averaged across all metrics, English again achieves the highest scores across datasets and models. Gemini leads in human evaluations, followed by GPT-4o, while MuLan models perform poorly in Hindi. VIST continues to outperform FlintstonesSV for English and Hindi, whereas Chinese results are less consistent. In the MLLM-as-Jury evaluation, MuLan SDXL outperforms the other two open-source models. While English shows higher results than Hindi and Chinese, the VIST dataset yields higher scores than FlintstonesSV across all three languages. The MLLM-as-Jury ablation study results are provided in Appendix Section~\ref{MLLM-as-Jury Ablation Study}.

%Table~\ref{tab:Average_Results_Across_Metrics} shows that when averaged across all metrics, English again achieves the highest scores across datasets and models. Gemini leads in human evaluations, followed by GPT-4o, while MuLan models perform poorly in Hindi. VIST continues to outperform FlintstonesSV for English and Hindi, whereas Chinese results are less consistent. In MLLM-as-Jury evaluation, Mulan SDXL outperforms the other two open-source models. While ENglish shows highest results than Hindi and Chinese, Vist Datasest indicates high results than FlintstonesSV in all three languages. MLLM-as-Jury ablation study results are provided in Appendix Section \ref{MLLM-as-Jury Ablation Study}\\

\renewcommand{\arraystretch}{1}
\begin{table}[h]
  \centering
  \resizebox{0.47\textwidth}{!}{%
  \begin{tabular}{lcccccc}
    \toprule
    \textbf{Models} 
      & \multicolumn{3}{c}{\textbf{FlintstonesSV}} 
      & \multicolumn{3}{c}{\textbf{VIST}} \\
    \cmidrule(lr){2-4}\cmidrule(lr){5-7}
      & \textbf{English} & \textbf{Hindi} & \textbf{Chinese} 
      & \textbf{English} & \textbf{Hindi} & \textbf{Chinese}  \\
    \midrule
    \rowcolor{gray!15}
    \multicolumn{7}{c}{\textbf{Human Evaluation $\uparrow$}} \\
    MuLan SD2v1       & 2.88 & 2.35 & 2.72 & 3.28 & 2.58 & 2.82 \\
    MuLan SD1v5       & 2.85 & 2.50 & 2.73 & 3.12 & 2.76 & 2.51 \\
    MuLan SDXL        & 3.01 & 2.22 & 3.03 & 3.16 & 2.70 & 2.95 \\
    \rowcolor{orange!15}
    GPT-4o            & 3.74 & 2.89 & 3.35 & 4.00 & 3.99 & 3.88 \\
    \rowcolor{orange!15}
    Gemini Pro 2.5    & \textbf{4.33} & \textbf{3.53} & \textbf{3.64} & \textbf{4.19} & \textbf{4.18} & \textbf{4.22} \\
    \addlinespace
    \rowcolor{gray!15}
    \multicolumn{7}{c}{\textbf{MLLM-as-Jury Evaluation $\uparrow$}} \\
    MuLan SD2v1       & 3.97 & 3.52 & 3.72 & 4.17 & 3.85 & 3.87 \\
    MuLan SD1v5       & 4.01 & 3.58 & 3.73 & 4.23 & 3.92 & 4.02 \\
    \rowcolor{orange!15}
    MuLan SDXL        & \textbf{4.08} & \textbf{3.69} & \textbf{3.84} & \textbf{4.32} & \textbf{4.06} & \textbf{4.12} \\
    \bottomrule
  \end{tabular}
  }
  \caption{Average scores across metrics from Human Evaluation (top) and MLLM-as-Jury Evaluation (bottom), spanning models, languages and datasets with score range (1-5).}
  \label{tab:Average_Results_Across_Metrics}
\end{table}

%Table~\ref{tab:Average_Results_Across_Metrics} presents the average evaluation scores from both human judges and the MLLM-as-Jury framework. English consistently receives the highest scores across both datasets and evaluation methods, indicating better model performance in that language. Gemini achieves the highest human scores in all settings, followed closely by GPT-4o. In contrast, the open-source MuLan models perform significantly worse in other languages, particularly in Hindi. 

%Across datasets, VIST generally yields higher scores than FlintstonesSV, especially in English and Hindi, suggesting better model performance on real-world images. However, Chinese scores fluctuate, showing inconsistency across settings. This pattern holds for both human and MLLM-as-Jury evaluations. Notably, MLLM-as-Jury scores are consistently and significantly higher than human scores, indicating the automatic evaluation tends to overestimate performance, especially for open-source models.

\begin{figure}[ht]
  \centering
  \includegraphics[width=0.48\textwidth]{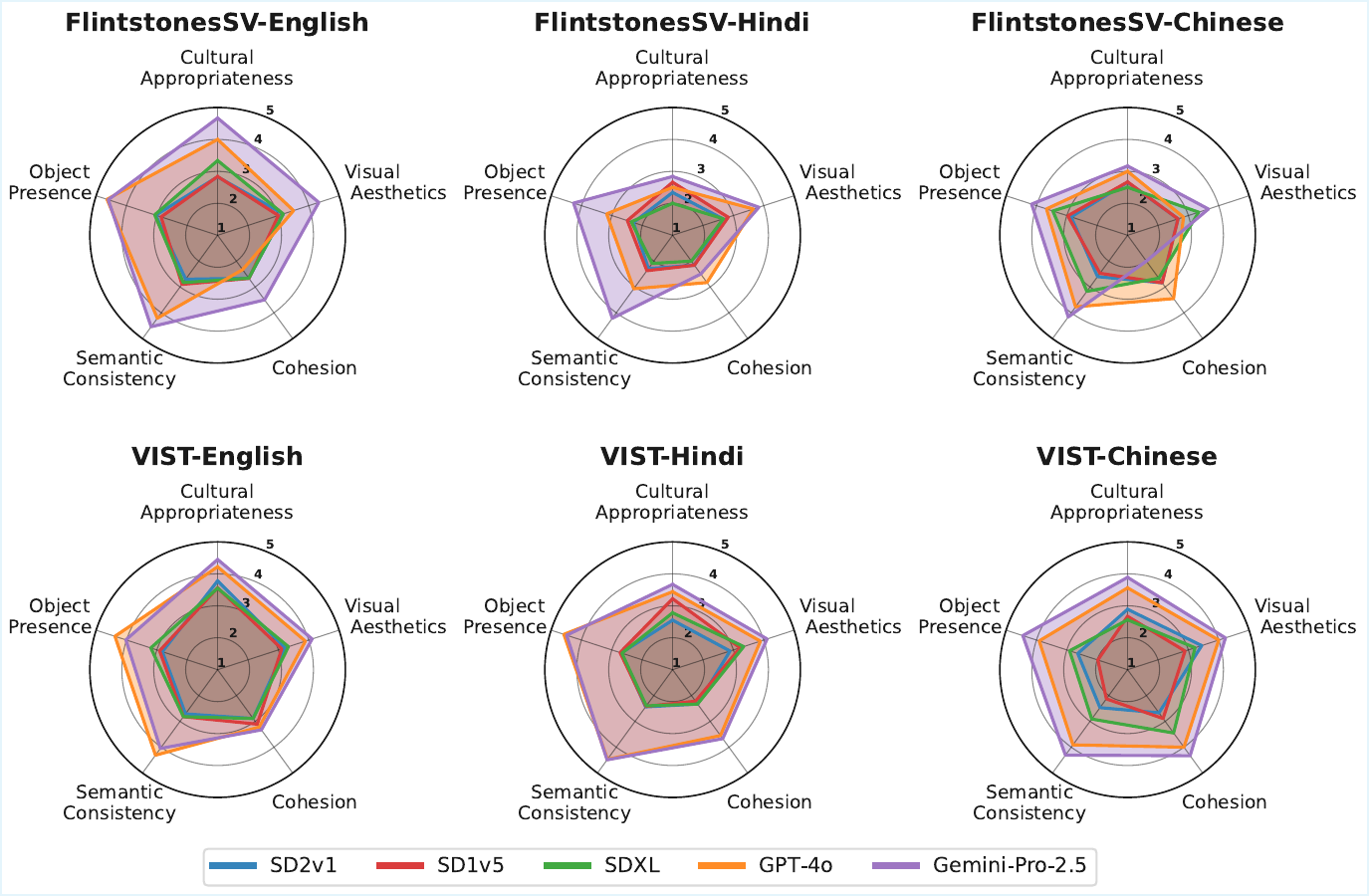}
  \caption{Human evaluation results across datasets, languages and evaluation metrics.}
  \label{fig:spiralplot}
\end{figure}

\noindent \textbf{Human Evaluation Results:} Figure~\ref{fig:spiralplot} illustrates that English scores are high in FlintstonesSV, especially for GPT-4o and Gemini. Open-source models perform poorly, except SDXL in Cultural Appropriateness. Hindi scores drop, with only Gemini scoring well in Object Presence and Semantic Consistency. Chinese results vary: Gemini scores low in Cohesion, where GPT-4o leads. In VIST, similar trends hold: English scores highest, Hindi improves slightly and Chinese results are better. GPT-4o and Gemini lead, while SD1v5 and SDXL show moderate performance. Closed-source models perform best across five metrics, while VIST results outperform FlintstonesSV.

\noindent\textbf{CLIP/FID Metrics Results: }
As shown in Table~\ref{tab:CLIP_and_FID_results} (see Appendix), CLIP and FID scores are reported for story images generated by three open-source multilingual text-to-image models. FlintstonesSV outputs achieve higher CLIP scores than VIST, indicating better alignment with scene descriptions due to its consistent generation of animated visual structure, while VIST’s open-ended real-world scenes result in more diverse generations. In contrast, FlintstonesSV shows higher FID distances from ground-truth animated scenes, likely due to stylistic and texture variations, whereas VIST images are closer in distribution to real-world references. Overall, English performs best across both metrics, followed by Hindi and Chinese for CLIP, and Chinese and Hindi for FID. The conflicting trends highlight that CLIP and FID capture different aspects of quality, making it difficult to rely solely on either metric for final evaluation.

%\noindent\textbf{Overall Average Results:} As shown in Table~\ref{tab:Average_Results_Across_Metrics}, Overall across all five metrics, English consistently receives the highest scores across datasets and models, reflecting better model performance. Gemini perform best in human scores, followed by GPT-4o. While MuLan models perform poor in Hindi. Scores on VIST dataset are generally higher than FlintstonesSV, particularly in English and Hindi while in Chinese scores fluctuate, showing inconsistency. MLLM-as-Jury scores follows the same behavior as Human scores but consistently higher than human scores, often overestimating performance for open-source models.\\

\begin{figure}[ht]
  \centering
  \includegraphics[width=0.47\textwidth]{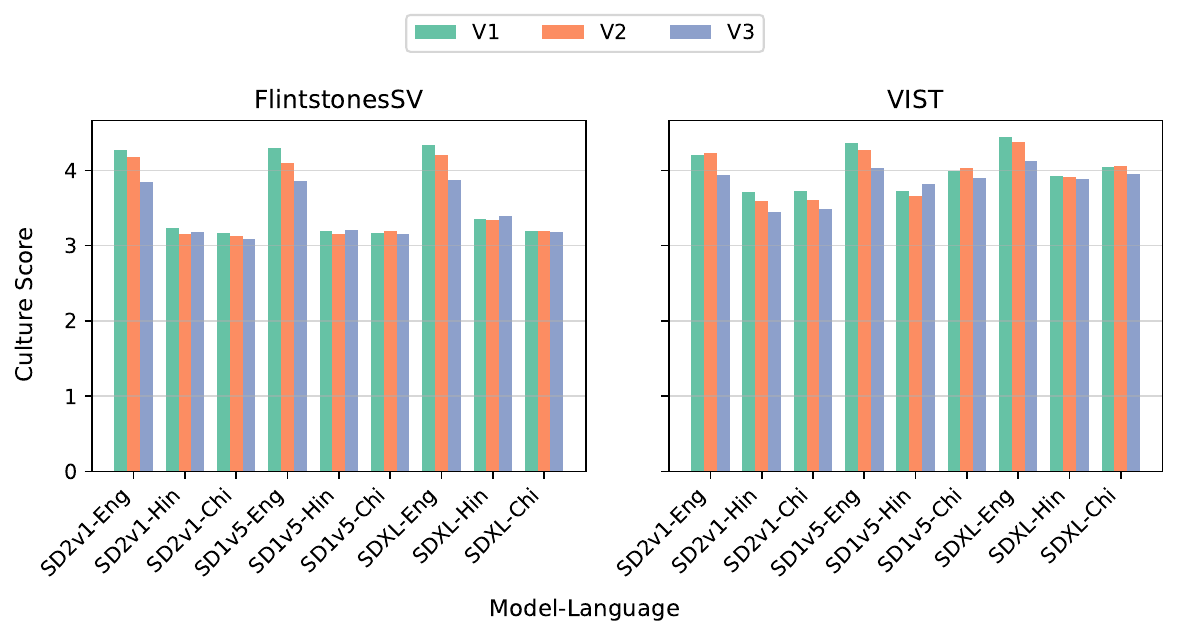}
  \caption{Progressive Culture evaluation scores by MLLM-as-Jury across three levels on FlintstonesSV and VIST datasets.}
  \label{fig:culturebar}
\end{figure}

\noindent \textbf{Progressive Multiculture Evaluation Results:} Figure~\ref{fig:culturebar} shows cultural appropriateness scores across three levels. For English, V1 scores highest over all models while decreasing its performance with V2 and V3, showing stricter judgments in evaluation with added focus points and illustrative examples. For Chinese and Hindi, scores remain stable between V1 to V3, indicating that the added structure has minimal impact because of the underlying Judge models' limited capability to evaluate cultural aspects in these languages by enriching prompts. Model SDXL outperforms SD2v1 and SD1v5 on both datasets, while overall scores high on VIST compared to FlintstonesSV, showing generated story adapts culture better in real-world dataset.

\begin{figure}[ht]
  \centering
  \includegraphics[width=0.48\textwidth]{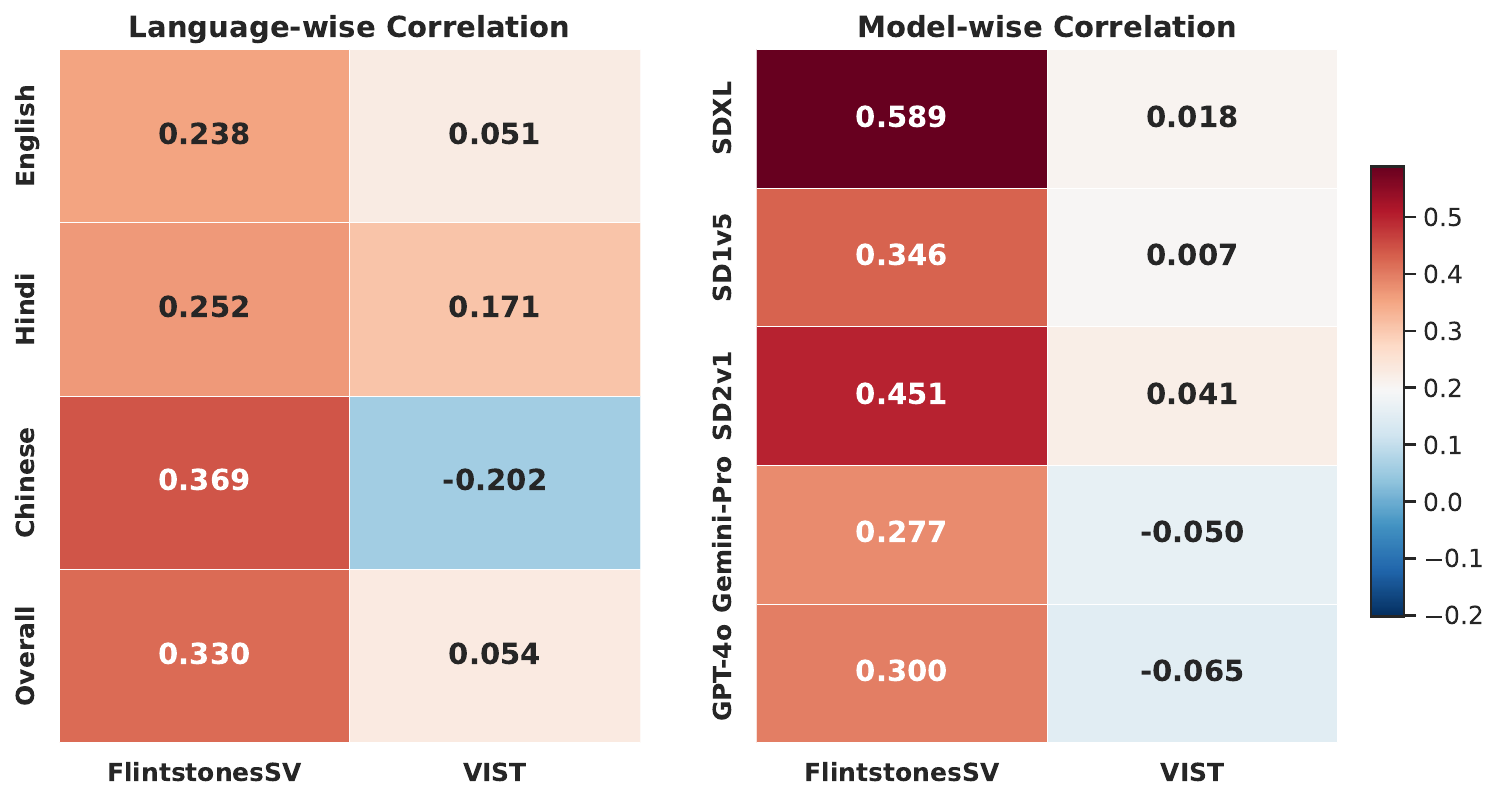}
  \caption{Correlation between Human and MLLM-as-Jury evaluation on FlintstonesSV and VIST datasets}
  \label{fig:Human_vs_Jury_Correlation_Scores}
\end{figure}

\noindent \textbf{Correlation between Human and MLLM-as-Jury Evaluation:} Figure~\ref{fig:Human_vs_Jury_Correlation_Scores} shows Spearman Rank Correlation between Human and MLLM-as-Jury evaluation. In FlintstonesSV, the MLLM-as-Jury evaluation aligns moderately with human evaluation across all languages and models. This is because FlintstonesSV features cartoon-based, fixed characters and repetitive story settings (e.g., cave, car, room, backyard). In contrast, VIST exhibits a lower correlation across languages and models, as it contains open-ended stories where scenes vary widely, and each story introduces new people, places, activities, ceremonies, and relationships. As shown in Figure~\ref{fig:Rating_Distribution_Plots} (see Appendix), the rating distribution of MLLM-as-Jury on VIST tends to produce higher scores compared to FlintstonesSV, where ratings are more dispersed and better aligned with human evaluations. The lower variance in MLLM-as-Jury ratings on VIST leads to a reduced Spearman correlation with human assessments. Overall, this analysis indicates that while MLLM-as-Jury can approximate human judgment effectively in simple, structured tasks, it tends to overestimate scores in complex or open-ended scenarios, making it a dataset-dependent automatic evaluation method.

\section{Discussion}

In this section, we address the two research questions focusing visualisation of the same story in different languages and cultural adaptation.

\begin{figure}[ht]
  \centering
  \includegraphics[width=0.48\textwidth]{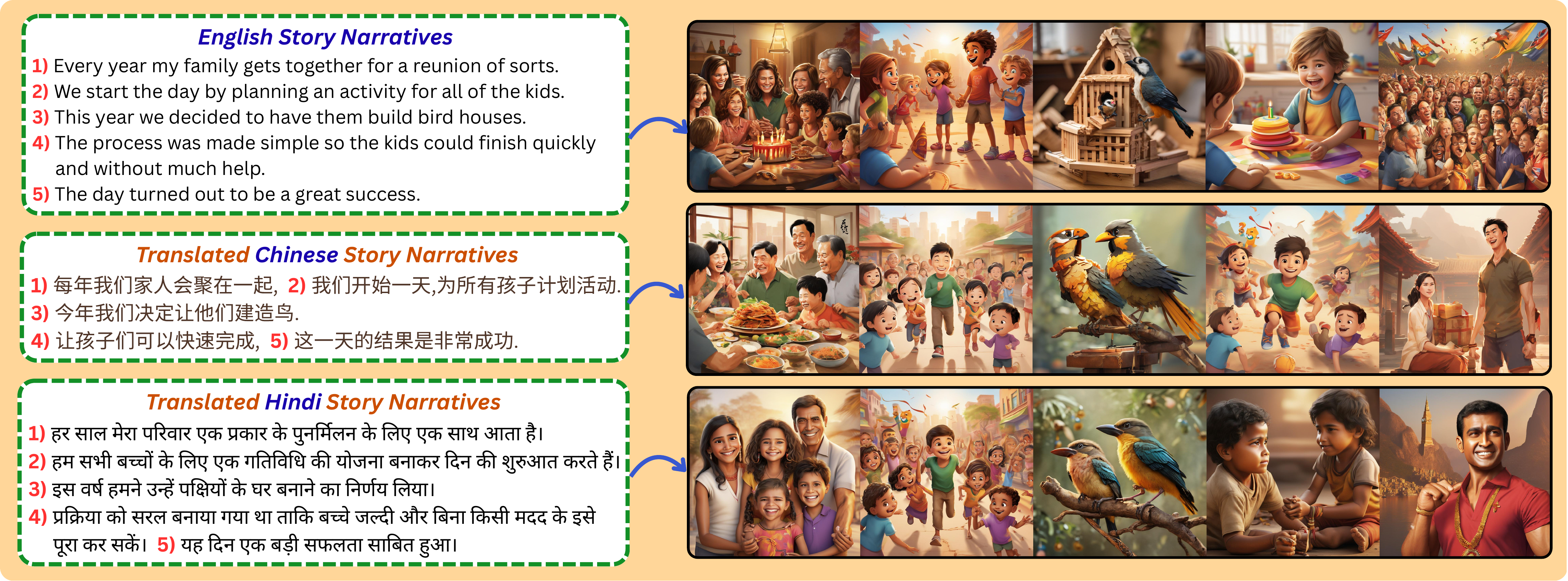}
  \caption{Visualization of the same story in different languages generated by \texttt{MuLan-SDXL} model.}
  \label{fig:story_in_3_language}
\end{figure}

\textbf{RQ1: How does the visualization of the same story vary across different languages?} As we can see from Figure \ref{fig:story_in_3_language},  
the generated image sequences across different languages reveal various cultural elements including hairstyles, facial features and apparel. In scene 1, for a family get-together, in Chinese, the food appears as traditional Chinese cuisine, whereas in the English setting, a cake-cutting and in Hindi, it is a casual get-together. These examples show that the same model generates culturally different visuals across different languages. Additionally, the concept of “success” is represented differently across languages: in Hindi, it is shown as an individual achievement, in Chinese as a joint success with a partner, while in English, a collective family success. 

Figure \ref{fig:same_story_across_different_models} (see Appendix) compares visualizations of the same story generated by different models in the same language. For the Chinese story on the left (a visit to a craft fair from VIST), open models such as SD1v5, SDXL, and SD2v1 retain strong cultural cues such as traditional clothing, facial features, food items, shop signs, and calligraphy. GPT-4o preserves some Chinese elements only in Scenes 3 and 5, and Gemini-Pro does so mainly in Scenes 2 and 5. The remaining scenes resemble Western settings. For the Hindi story on the right (Wilma and Fred talking indoors from FlintstoneSV), only SDXL introduces partial Indian cues, though even there the characters and environments remain largely Westernized. Across the other models, traditional Indian attire, features, and household settings are mostly absent, showing a clear bias toward Western visual culture.

Table \ref{tab:MLLM_as_Jury_results} and Figure \ref{fig:spiralplot} show that models generate more culturally appropriate visual stories for real-world narratives in VIST than for animated stories in FlintstonesSV. VIST’s open-ended structure gives models flexibility to adapt cultural cues across languages, whereas FlintstonesSV is anchored to Western character identities (Fred, Wilma, Betty, Barney). As also seen in Figure \ref{fig:same_story_across_different_models} for Hindi outputs, these fixed references restrict the models’ ability to produce culturally aligned visuals, making animated datasets inherently less adaptable than real-world ones.

\begin{figure}[ht]
  \centering
  \includegraphics[width=0.48\textwidth]{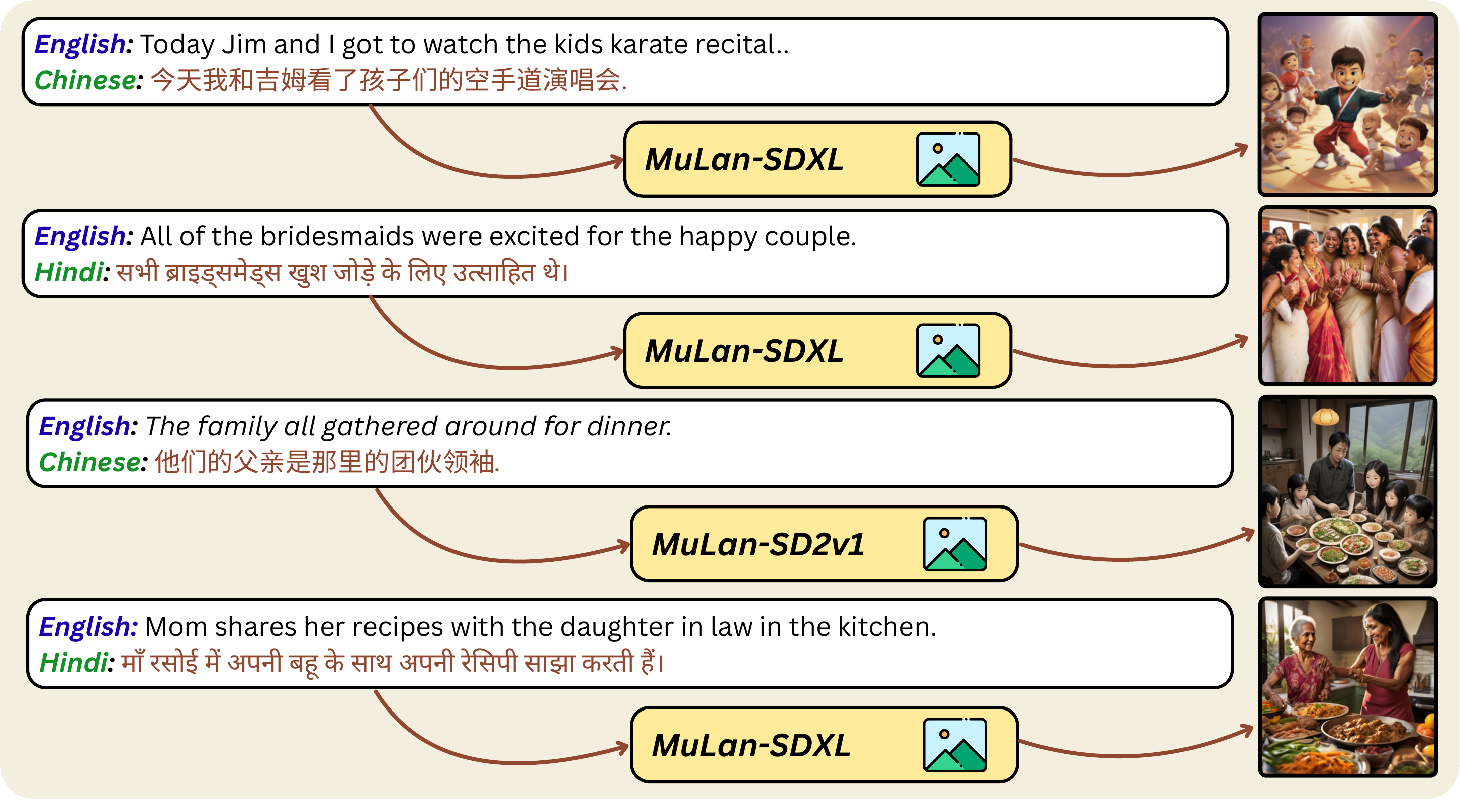}
  \caption{Model outputs demonstrating cultural adaptation}
  \label{fig:culture_adaptation_VIST}
\end{figure}

% \textbf{RQ4: Effect of open and closed source models on multicultural story visualization.} Our human evaluation results shown in Table \ref{tab:Average_Results_Across_Metrics}, reveals that closed-source models consistently outperform open-source models across all story aspects. Among them, Gemini-Flash 2.5 demonstrates superior performance compared to GPT-4o. Closed-source models also exhibit stronger cultural adaptation, particularly in English, followed by Chinese and Hindi across both datasets. This analysis highlights the limitations of open-source multilingual T2I models in accurately following prompt instructions and generating culturally faithful stories.

\textbf{RQ2: How does story visualization adapt to different cultures?} From our multicultural analysis, we found that current models generate Western cultural elements most accurately, while their adaptation of Chinese and Hindi cultures is moderate and low, respectively. For example, in Figure \ref{fig:culture_adaptation_VIST}, \textbf{Ex1:} the story scene in Chinese about karate adapts to Chinese karate clothes. \textbf{Ex2:} The story scene in Hindi about a wedding, where the scene adapts to an Indian wedding with culturally relevant 'saari' dresses for the bride and bridesmaids. \textbf{Ex3:} The story scene in Chinese about a family dinner, where the scene adapts to Chinese food with a traditional light bulb. \textbf{Ex4:} The story scene in Hindi about a mother sharing a recipe with her daughter-in-law in the kitchen. The scene adapts to Indian food, steel utensils, and Indian clothes. These examples show a minimal level of cultural adaptation by default in multilingual T2I models. This analysis further underscores the need to develop more culturally grounded story visualization models.

\section{Culture Error Analysis}
We identified several errors during the evaluation of cultural aspects in the generated stories. Figure~\ref{fig:Hindi_Error_analysis_Sample} (see Appendix) shows a Hindi story depicting a family vacation with friends. In Scene 1, the absence of footwear, while Scene 4 exhibits Western attire, shows a misrepresentation of Indian culture. Similarly, Figure~\ref{fig:English_Error_analysis_Sample} (see Appendix) presents an example from a Chinese story about families enjoying a game together. In Scene 2, the flag closely resembles the American flag. In Scene 4, the male character’s exaggerated height over the woman and the girl reflects a stereotypical “dominant male” portrayal, which is inconsistent with typical Chinese family depictions. Scene 5 further displays characters with narrowed eyes and nearly identical facial expressions, signalling culturally inappropriate and stereotyped visuals.

\section{Conclusion}
We studied the multicultural analysis of story visualization on real-world and animated datasets. Our study reveals that current multilingual T2I models exhibit a strong bias towards Western culture. Human and MLLM-as-Jury evaluation show that models generate high culturally appropriate stories in English, moderate in Chinese and low for Hindi. More specifically, on a real-world story dataset, models generate better culturally relevant stories compared to an animated dataset. Our Progressive Cultural Evaluation framework shows that providing cultural focus points and illustrative examples helps to evaluate detailed cultural aspects in English. However, for Chinese and Hindi, not much benefit is observed, reflecting the underlying Judge models' limited ability to accurately identify cultural aspects in different languages. MLLM-as-Jury evaluation results approximate human judgment with higher correlation on the FlintstonesSV dataset than on the open-ended story dataset VIST. These findings highlight the need to develop culturally grounded story visualization approaches that faithfully represent diverse cultures and languages.

% These findings highlight the need to develop culturally grounded approaches for story visualization that faithfully represent diverse cultures and languages.

% while on Chinese and Hindi it shows no benefit as limited capability of underlying Judge models to identify correct cultural aspects. 

\section*{Limitations}
This work has several limitations. \textbf{1)} For multilingual story visualization, we generate each scene independently using multilingual T2I models, as our primary focus is on the cultural aspects of each scene rather than the narrative coherence across scenes. \textbf{2)} The scope of multicultural analysis in this study is limited to three languages: English, Chinese and Hindi. We plan to expand this analysis in future work by incorporating a more diverse set of languages and cultures. \textbf{3)} While this work focuses on analyzing cultural aspects, our future efforts will address these discrepancies by developing culturally grounded methods that adapt generated stories to the target culture.

% \textbf{2)} In our MLLM-as-Jury evaluation, we observe that the aggregated scores from the three judge models tend to be consistently higher than human evaluations, indicating a bias towards overrating responses.

\section*{Acknowledgments}
This research work is funded by Research Ireland under Grant Number SFI/12/RC/2289\_P2 (Insight), co-funded by the European Regional Development Fund.

% This publication has emanated from research conducted with the financial support of Research Ireland under Grant Number 12/RC/2289_P2 - Insight
% Research Ireland Centre for Data Analytics. For
% the purpose of Open Access, the author has applied
% a CC BY public copyright licence to any Author
% Accepted Manuscript version arising from this submission.

\bibliography{custom}

@InProceedings{Pan_2024_WACV,
    author    = {Pan, Xichen and Qin, Pengda and Li, Yuhong and Xue, Hui and Chen, Wenhu},
    title     = {Synthesizing Coherent Story With Auto-Regressive Latent Diffusion Models},
    booktitle = {Proceedings of the IEEE/CVF Winter Conference on Applications of Computer Vision (WACV)},
    month     = {January},
    year      = {2024},
    pages     = {2920-2930}
}

@inproceedings{rahman2023make,
  title={Make-a-story: Visual memory conditioned consistent story generation},
  author={Rahman, Tanzila and Lee, Hsin-Ying and Ren, Jian and Tulyakov, Sergey and Mahajan, Shweta and Sigal, Leonid},
  booktitle={Proceedings of the IEEE/CVF Conference on Computer Vision and Pattern Recognition},
  pages={2493--2502},
  year={2023}
}

@article{shen2023large,
  title={Large Language Models as Consistent Story Visualizers},
  author={Shen, Xiaoqian and Elhoseiny, Mohamed},
  journal={arXiv preprint arXiv:2312.02252},
  year={2023}
}

@article{zheng2024temporalstory,
  title={Temporalstory: Enhancing consistency in story visualization using spatial-temporal attention},
  author={Zheng, Sixiao and Fu, Yanwei},
  journal={arXiv e-prints},
  pages={arXiv--2407},
  year={2024}
}

@article{baldridge2024imagen,
  title={Imagen 3},
  author={Baldridge, Jason and Bauer, Jakob and Bhutani, Mukul and Brichtova, Nicole and Bunner, Andrew and Chan, Kelvin and Chen, Yichang and Dieleman, Sander and Du, Yuqing and Eaton-Rosen, Zach and others},
  journal={arXiv preprint arXiv:2408.07009},
  year={2024}
}

@misc{gupta2018imaginethisscriptscompositions,
      title={Imagine This! Scripts to Compositions to Videos}, 
      author={Tanmay Gupta and Dustin Schwenk and Ali Farhadi and Derek Hoiem and Aniruddha Kembhavi},
      year={2018},
      eprint={1804.03608},
      archivePrefix={arXiv},
      primaryClass={cs.CV},
      url={https://arxiv.org/abs/1804.03608}, 
}

@article{betker2023improving,
  title={Improving image generation with better captions},
  author={Betker, James and Goh, Gabriel and Jing, Li and Brooks, Tim and Wang, Jianfeng and Li, Linjie and Ouyang, Long and Zhuang, Juntang and Lee, Joyce and Guo, Yufei and others},
  journal={Computer Science. https://cdn. openai. com/papers/dall-e-3. pdf},
  volume={2},
  number={3},
  pages={8},
  year={2023}
}

@article{podell2023sdxl,
  title={Sdxl: Improving latent diffusion models for high-resolution image synthesis},
  author={Podell, Dustin and English, Zion and Lacey, Kyle and Blattmann, Andreas and Dockhorn, Tim and M{\"u}ller, Jonas and Penna, Joe and Rombach, Robin},
  journal={arXiv preprint arXiv:2307.01952},
  year={2023}
}

@inproceedings{zhang2023adding,
  title={Adding conditional control to text-to-image diffusion models},
  author={Zhang, Lvmin and Rao, Anyi and Agrawala, Maneesh},
  booktitle={Proceedings of the IEEE/CVF international conference on computer vision},
  pages={3836--3847},
  year={2023}
}

@misc{labs2025flux1kontextflowmatching,
      title={FLUX.1 Kontext: Flow Matching for In-Context Image Generation and Editing in Latent Space},
      author={Black Forest Labs and Stephen Batifol and Andreas Blattmann and Frederic Boesel and Saksham Consul and Cyril Diagne and Tim Dockhorn and Jack English and Zion English and Patrick Esser and Sumith Kulal and Kyle Lacey and Yam Levi and Cheng Li and Dominik Lorenz and Jonas Müller and Dustin Podell and Robin Rombach and Harry Saini and Axel Sauer and Luke Smith},
      year={2025},
      eprint={2506.15742},
      archivePrefix={arXiv},
      primaryClass={cs.GR},
      url={https://arxiv.org/abs/2506.15742},
}

@article{blattmann2023stable,
  title={Stable video diffusion: Scaling latent video diffusion models to large datasets},
  author={Blattmann, Andreas and Dockhorn, Tim and Kulal, Sumith and Mendelevitch, Daniel and Kilian, Maciej and Lorenz, Dominik and Levi, Yam and English, Zion and Voleti, Vikram and Letts, Adam and others},
  journal={arXiv preprint arXiv:2311.15127},
  year={2023}
}

@article{videoworldsimulators2024,
  title={Video generation models as world simulators},
  author={Tim Brooks and Bill Peebles and Connor Holmes and Will DePue and Yufei Guo and Li Jing and David Schnurr and Joe Taylor and Troy Luhman and Eric Luhman and Clarence Ng and Ricky Wang and Aditya Ramesh},
  year={2024},
  url={https://openai.com/research/video-generation-models-as-world-simulators},
}

@article{veo2024,
  title={Veo},
  author={Abhishek Sharma and Adams Yu and Ali Razavi and Andeep Toor and Andrew Pierson and Ankush Gupta and Austin Waters and Aäron van den Oord and Daniel Tanis and Dumitru Erhan and Eric Lau and Eleni Shaw and Gabe Barth-Maron and Greg Shaw and Han Zhang and Henna Nandwani and Hernan Moraldo and Hyunjik Kim and Irina Blok and Jakob Bauer and Jeff Donahue and Junyoung Chung and Kory Mathewson and Kurtis David and Lasse Espeholt and Marc van Zee and Matt McGill and Medhini Narasimhan and Miaosen Wang and Mikołaj Bińkowski and Mohammad Babaeizadeh and Mohammad Taghi Saffar and Nando de Freitas and Nick Pezzotti and Pieter-Jan Kindermans and Poorva Rane and Rachel Hornung and Robert Riachi and Ruben Villegas and Rui Qian and Sander Dieleman and Serena Zhang and Serkan Cabi and Shixin Luo and Shlomi Fruchter and Signe Nørly and Srivatsan Srinivasan and Tobias Pfaff and Tom Hume and Vikas Verma and Weizhe Hua and William Zhu and Xinchen Yan and Xinyu Wang and Yelin Kim and Yuqing Du and Yutian Chen},
  url={https://deepmind.google/technologies/veo/},
  year={2024}
}

@inproceedings{huang2016visual,
  title={Visual Storytelling},
  author={Huang, Ting-Hao K. and Ferraro, Francis and Mostafazadeh, Nasrin and Misra, Ishan and Devlin, Jacob and Agrawal, Aishwarya and Girshick, Ross and He, Xiaodong and Kohli, Pushmeet and Batra, Dhruv and others},
  booktitle={15th Annual Conference of the North American Chapter of the Association for Computational Linguistics (NAACL 2016)},
  year={2016}
}

@article{yang2024seed,
  title={Seed-story: Multimodal long story generation with large language model},
  author={Yang, Shuai and Ge, Yuying and Li, Yang and Chen, Yukang and Ge, Yixiao and Shan, Ying and Chen, Yingcong},
  journal={arXiv preprint arXiv:2407.08683},
  year={2024}
}

@article{zhou2024storydiffusion,
  title={Storydiffusion: Consistent self-attention for long-range image and video generation},
  author={Zhou, Yupeng and Zhou, Daquan and Cheng, Ming-Ming and Feng, Jiashi and Hou, Qibin},
  journal={Advances in Neural Information Processing Systems},
  volume={37},
  pages={110315--110340},
  year={2024}
}

@article{mao2024story,
  title={Story-Adapter: A Training-free Iterative Framework for Long Story Visualization},
  author={Mao, Jiawei and Huang, Xiaoke and Xie, Yunfei and Chang, Yuanqi and Hui, Mude and Xu, Bingjie and Zhou, Yuyin},
  journal={CoRR},
  year={2024}
}

@inproceedings{radford2021learning,
  title={Learning transferable visual models from natural language supervision},
  author={Radford, Alec and Kim, Jong Wook and Hallacy, Chris and Ramesh, Aditya and Goh, Gabriel and Agarwal, Sandhini and Sastry, Girish and Askell, Amanda and Mishkin, Pamela and Clark, Jack and others},
  booktitle={International conference on machine learning},
  pages={8748--8763},
  year={2021},
  organization={PMLR}
}

@article{heusel2017gans,
  title={Gans trained by a two time-scale update rule converge to a local nash equilibrium},
  author={Heusel, Martin and Ramsauer, Hubert and Unterthiner, Thomas and Nessler, Bernhard and Hochreiter, Sepp},
  journal={Advances in neural information processing systems},
  volume={30},
  year={2017}
}

@inproceedings{lee2024prometheus,
  title={Prometheus-vision: Vision-language model as a judge for fine-grained evaluation},
  author={Lee, Seongyun and Kim, Seungone and Park, Sue and Kim, Geewook and Seo, Minjoon},
  booktitle={Findings of the Association for Computational Linguistics ACL 2024},
  pages={11286--11315},
  year={2024}
}

@inproceedings{chen2024mllm,
  title={MLLM-as-a-Judge: assessing multimodal LLM-as-a-Judge with vision-language benchmark},
  author={Chen, Dongping and Chen, Ruoxi and Zhang, Shilin and Wang, Yaochen and Liu, Yinuo and Zhou, Huichi and Zhang, Qihui and Wan, Yao and Zhou, Pan and Sun, Lichao},
  booktitle={Proceedings of the 41st International Conference on Machine Learning},
  pages={6562--6595},
  year={2024}
}

@inproceedings{kocmi-federmann-2023-large,
    title = "Large Language Models Are State-of-the-Art Evaluators of Translation Quality",
    author = "Kocmi, Tom  and
      Federmann, Christian",
    editor = "Nurminen, Mary  and
      Brenner, Judith  and
      Koponen, Maarit  and
      Latomaa, Sirkku  and
      Mikhailov, Mikhail  and
      Schierl, Frederike  and
      Ranasinghe, Tharindu  and
      Vanmassenhove, Eva  and
      Vidal, Sergi Alvarez  and
      Aranberri, Nora  and
      Nunziatini, Mara  and
      Escart{\'i}n, Carla Parra  and
      Forcada, Mikel  and
      Popovic, Maja  and
      Scarton, Carolina  and
      Moniz, Helena",
    booktitle = "Proceedings of the 24th Annual Conference of the European Association for Machine Translation",
    month = jun,
    year = "2023",
    address = "Tampere, Finland",
    publisher = "European Association for Machine Translation",
    url = "https://aclanthology.org/2023.eamt-1.19/",
    pages = "193--203"
}

@inproceedings{chiang2023can,
  title={Can Large Language Models Be an Alternative to Human Evaluations?},
  author={Chiang, Cheng-Han and Lee, Hung-Yi},
  booktitle={Proceedings of the 61st Annual Meeting of the Association for Computational Linguistics (Volume 1: Long Papers)},
  pages={15607--15631},
  year={2023}
}

@article{zhu2023judgelm,
  title={Judgelm: Fine-tuned large language models are scalable judges},
  author={Zhu, Lianghui and Wang, Xinggang and Wang, Xinlong},
  journal={arXiv preprint arXiv:2310.17631},
  year={2023}
}

@article{xing2024mulan,
  title={Mulan: Adapting multilingual diffusion models for hundreds of languages with negligible cost},
  author={Xing, Sen and Zhong, Muyan and Lai, Zeqiang and Li, Liangchen and Liu, Jiawen and Wang, Yaohui and Dai, Jifeng and Wang, Wenhai},
  journal={arXiv preprint arXiv:2412.01271},
  year={2024}
}

@article{bai2025qwen2,
  title={Qwen2. 5-vl technical report},
  author={Bai, Shuai and Chen, Keqin and Liu, Xuejing and Wang, Jialin and Ge, Wenbin and Song, Sibo and Dang, Kai and Wang, Peng and Wang, Shijie and Tang, Jun and others},
  journal={arXiv preprint arXiv:2502.13923},
  year={2025}
}

@article{alam2024maya,
  title={Maya: An Instruction Finetuned Multilingual Multimodal Model},
  author={Alam, Nahid and Kanjula, Karthik Reddy and Guthikonda, Surya and Chung, Timothy and Vegesna, Bala Krishna S and Das, Abhipsha and Susevski, Anthony and Chan, Ryan Sze-Yin and Uddin, SM and Islam, Shayekh Bin and others},
  journal={arXiv preprint arXiv:2412.07112},
  year={2024}
}

@inproceedings{yadav2025cultural,
author = {Yadav, Srishti and Zhang, Zhi and Hershcovich, Daniel and Shutova, Ekaterina},
year = {2025},
month = {01},
pages = {7592-7608},
title = {Beyond Words: Exploring Cultural Value Sensitivity in Multimodal Models},
doi = {10.18653/v1/2025.findings-naacl.422}
}

@inproceedings{khanuja-etal-2024-image,
    title = "An image speaks a thousand words, but can everyone listen? On image transcreation for cultural relevance",
    author = "Khanuja, Simran  and
      Ramamoorthy, Sathyanarayanan  and
      Song, Yueqi  and
      Neubig, Graham",
    editor = "Al-Onaizan, Yaser  and
      Bansal, Mohit  and
      Chen, Yun-Nung",
    booktitle = "Proceedings of the 2024 Conference on Empirical Methods in Natural Language Processing",
    month = nov,
    year = "2024",
    address = "Miami, Florida, USA",
    publisher = "Association for Computational Linguistics",
    url = "https://aclanthology.org/2024.emnlp-main.573/",
    doi = "10.18653/v1/2024.emnlp-main.573",
    pages = "10258--10279"
}

@misc{bhatia2024multicultural,
      title={From Local Concepts to Universals: Evaluating the Multicultural Understanding of Vision-Language Models}, 
      author={Mehar Bhatia and Sahithya Ravi and Aditya Chinchure and Eunjeong Hwang and Vered Shwartz},
      year={2024},
      eprint={2407.00263},
      archivePrefix={arXiv},
      primaryClass={cs.CL},
      url={https://arxiv.org/abs/2407.00263}, 
}

@misc{bayramli2025cultdiff,
      title={Diffusion Models Through a Global Lens: Are They Culturally Inclusive?}, 
      author={Zahra Bayramli and Ayhan Suleymanzade and Na Min An and Huzama Ahmad and Eunsu Kim and Junyeong Park and James Thorne and Alice Oh},
      year={2025},
      eprint={2502.08914},
      archivePrefix={arXiv},
      primaryClass={cs.CV},
      url={https://arxiv.org/abs/2502.08914}, 
}

@misc{romero2024cvqa,
      title={CVQA: Culturally-diverse Multilingual Visual Question Answering Benchmark}, 
      author={David Romero and Chenyang Lyu and Haryo Akbarianto Wibowo and Teresa Lynn and Injy Hamed and Aditya Nanda Kishore and Aishik Mandal and Alina Dragonetti and Artem Abzaliev and Atnafu Lambebo Tonja and Bontu Fufa Balcha and Chenxi Whitehouse and Christian Salamea and Dan John Velasco and David Ifeoluwa Adelani and David Le Meur and Emilio Villa-Cueva and Fajri Koto and Fauzan Farooqui and Frederico Belcavello and Ganzorig Batnasan and Gisela Vallejo and Grainne Caulfield and Guido Ivetta and Haiyue Song and Henok Biadglign Ademtew and Hernán Maina and Holy Lovenia and Israel Abebe Azime and Jan Christian Blaise Cruz and Jay Gala and Jiahui Geng and Jesus-German Ortiz-Barajas and Jinheon Baek and Jocelyn Dunstan and Laura Alonso Alemany and Kumaranage Ravindu Yasas Nagasinghe and Luciana Benotti and Luis Fernando D'Haro and Marcelo Viridiano and Marcos Estecha-Garitagoitia and Maria Camila Buitrago Cabrera and Mario Rodríguez-Cantelar and Mélanie Jouitteau and Mihail Mihaylov and Mohamed Fazli Mohamed Imam and Muhammad Farid Adilazuarda and Munkhjargal Gochoo and Munkh-Erdene Otgonbold and Naome Etori and Olivier Niyomugisha and Paula Mónica Silva and Pranjal Chitale and Raj Dabre and Rendi Chevi and Ruochen Zhang and Ryandito Diandaru and Samuel Cahyawijaya and Santiago Góngora and Soyeong Jeong and Sukannya Purkayastha and Tatsuki Kuribayashi and Teresa Clifford and Thanmay Jayakumar and Tiago Timponi Torrent and Toqeer Ehsan and Vladimir Araujo and Yova Kementchedjhieva and Zara Burzo and Zheng Wei Lim and Zheng Xin Yong and Oana Ignat and Joan Nwatu and Rada Mihalcea and Thamar Solorio and Alham Fikri Aji},
      year={2024},
      eprint={2406.05967},
      archivePrefix={arXiv},
      primaryClass={cs.CV},
      url={https://arxiv.org/abs/2406.05967}, 
}

@misc{liu2025culturevlm,
      title={CultureVLM: Characterizing and Improving Cultural Understanding of Vision-Language Models for over 100 Countries}, 
      author={Shudong Liu and Yiqiao Jin and Cheng Li and Derek F. Wong and Qingsong Wen and Lichao Sun and Haipeng Chen and Xing Xie and Jindong Wang},
      year={2025},
      eprint={2501.01282},
      archivePrefix={arXiv},
      primaryClass={cs.AI},
      url={https://arxiv.org/abs/2501.01282}, 
}

@misc{adilazuarda2023culture,
      title={Towards Measuring and Modeling "Culture" in LLMs: A Survey}, 
      author={Muhammad Farid Adilazuarda and Sagnik Mukherjee and Pradhyumna Lavania and Siddhant Singh and Alham Fikri Aji and Jacki O'Neill and Ashutosh Modi and Monojit Choudhury},
      year={2024},
      eprint={2403.15412},
      archivePrefix={arXiv},
      primaryClass={cs.CY},
      url={https://arxiv.org/abs/2403.15412}, 
}

@misc{liu2023culturalgap,
      title={On the Cultural Gap in Text-to-Image Generation}, 
      author={Bingshuai Liu and Longyue Wang and Chenyang Lyu and Yong Zhang and Jinsong Su and Shuming Shi and Zhaopeng Tu},
      year={2023},
      eprint={2307.02971},
      archivePrefix={arXiv},
      primaryClass={cs.CV},
      url={https://arxiv.org/abs/2307.02971}, 
}

@misc{bhalerao2025mosaig,
      title={Multi-Agent Multimodal Models for Multicultural Text to Image Generation}, 
      author={Parth Bhalerao and Mounika Yalamarty and Brian Trinh and Oana Ignat},
      year={2025},
      eprint={2502.15972},
      archivePrefix={arXiv},
      primaryClass={cs.CV},
      url={https://arxiv.org/abs/2502.15972}, 
}

@inproceedings{nayak2024benchmarking,
  title={Benchmarking Vision Language Models for Cultural Understanding},
  author={Nayak, Shravan and Jain, Kanishk and Awal, Rabiul and Reddy, Siva and Steenkiste, Sjoerd and Hendricks, Lisa and Stanczak, Karolina and Agrawal, Aishwarya},
  booktitle={Proceedings of the 2024 Conference on Empirical Methods in Natural Language Processing},
  pages={5769--5790},
  year={2024}
}

@misc{ventura2023culturalpov,
      title={Navigating Cultural Chasms: Exploring and Unlocking the Cultural POV of Text-To-Image Models}, 
      author={Mor Ventura and Eyal Ben-David and Anna Korhonen and Roi Reichart},
      year={2024},
      eprint={2310.01929},
      archivePrefix={arXiv},
      primaryClass={cs.CL},
      url={https://arxiv.org/abs/2310.01929}, 
}

@article{bruner2010narrative,
  title={Narrative, culture, and mind},
  author={Bruner, Jerome},
  journal={Telling stories: Language, narrative, and social life},
  volume={46},
  pages={49},
  year={2010},
  publisher={Georgetown University Press Washington, DC}
}

@inproceedings{rei-etal-2020-comet,
    title = "{COMET}: A Neural Framework for {MT} Evaluation",
    author = "Rei, Ricardo  and
      Stewart, Craig  and
      Farinha, Ana C  and
      Lavie, Alon",
    editor = "Webber, Bonnie  and
      Cohn, Trevor  and
      He, Yulan  and
      Liu, Yang",
    booktitle = "Proceedings of the 2020 Conference on Empirical Methods in Natural Language Processing (EMNLP)",
    month = nov,
    year = "2020",
    address = "Online",
    publisher = "Association for Computational Linguistics",
    url = "https://aclanthology.org/2020.emnlp-main.213/",
    doi = "10.18653/v1/2020.emnlp-main.213",
    pages = "2685--2702"
}

@article{verga2024replacing,
  title={Replacing judges with juries: Evaluating llm generations with a panel of diverse models},
  author={Verga, Pat and Hofstatter, Sebastian and Althammer, Sophia and Su, Yixuan and Piktus, Aleksandra and Arkhangorodsky, Arkady and Xu, Minjie and White, Naomi and Lewis, Patrick},
  journal={arXiv preprint arXiv:2404.18796},
  year={2024}
}

@misc{nllbteam2022languageleftbehindscaling,
      title={No Language Left Behind: Scaling Human-Centered Machine Translation}, 
      author={NLLB Team and Marta R. Costa-jussà and James Cross and Onur Çelebi and Maha Elbayad and Kenneth Heafield and Kevin Heffernan and Elahe Kalbassi and Janice Lam and Daniel Licht and Jean Maillard and Anna Sun and Skyler Wang and Guillaume Wenzek and Al Youngblood and Bapi Akula and Loic Barrault and Gabriel Mejia Gonzalez and Prangthip Hansanti and John Hoffman and Semarley Jarrett and Kaushik Ram Sadagopan and Dirk Rowe and Shannon Spruit and Chau Tran and Pierre Andrews and Necip Fazil Ayan and Shruti Bhosale and Sergey Edunov and Angela Fan and Cynthia Gao and Vedanuj Goswami and Francisco Guzmán and Philipp Koehn and Alexandre Mourachko and Christophe Ropers and Safiyyah Saleem and Holger Schwenk and Jeff Wang},
      year={2022},
      eprint={2207.04672},
      archivePrefix={arXiv},
      primaryClass={cs.CL},
      url={https://arxiv.org/abs/2207.04672}, 
}

\appendix

\section{Appendix}
\label{sec:appendix}

This appendix details the translation details, human evaluation process with examples, culture error analysis, MLLM-as-Jury ablation study, Human and MLLM-as-Jury correlation details, experimental setup details, and prompts utilized within the MLLM-as-Jury evaluation framework.

\subsection{Reference Free Translation}
\label{ref_free_translation}

This section presents reference-free translation quality scores computed using COMET-QE\footnote{\url{https://github.com/Unbabel/COMET}}, a quality estimation model for machine translation. We evaluate English–Chinese and English–Hindi translations generated by the NLLB-200 model on both the VIST and FlintstonesSV datasets. COMET-QE allows us to estimate translation quality without requiring gold-standard reference translations. The scores reported here provide a reliable, model-based approximation of fluency and adequacy in the translated outputs, supporting systematic comparison across languages and datasets. As shown in Table \ref{tab:nllb_scores}, VIST consistently achieves higher COMET-QE scores than FlintstonesSV for both translation directions, with the average scores indicating slightly stronger performance for English–Chinese compared to English–Hindi.

\begin{table}[h]
    \centering
    \small
    \renewcommand{\arraystretch}{1.2}
    \begin{tabular}{@{}lcc@{}}
        \toprule
        \textbf{Dataset} & \textbf{Eng -> Hindi ( $\uparrow$)} & \textbf{Eng -> Chinese ($\uparrow$)} \\
        \midrule
        FlintstonesSV & 0.4537 & 0.3902 \\
        VIST           & 0.8224 & 0.7432 \\
        \rowcolor{blue!15}
        \textbf{Average} & \textbf{0.5667} & \textbf{0.6380} \\
        \bottomrule
    \end{tabular}
    \caption{Reference-free COMET-QE translation quality scores for the NLLB-200 model on the VIST and FlintstonesSV datasets for English–Hindi and English–Chinese translation.}
    \label{tab:nllb_scores}
\end{table}

\subsection{Human Evaluation}
\label{human_eval_appendix}
We conducted a human evaluation on random samples. 25 test samples spanned over 5 models and 2 datasets with three evaluators per language: English, Hindi and Chinese. The evaluation focused on five key metrics: \texttt{Cultural Appropriateness}, \texttt{Visual Aesthetics}, \texttt{Cohesion}, \texttt{Semantic Consistency} and \texttt{Object Presence}. Given the subjective nature of these metrics, inter-rater agreement was measured by weighted Cohen’s 
$\kappa$ scores shown in Table~\ref{tab:kappa_scores} were generally low to moderate across languages, which is typical for nuanced human judgment tasks.

\begin{figure}[ht]
  \centering
  \includegraphics[width=0.48\textwidth]{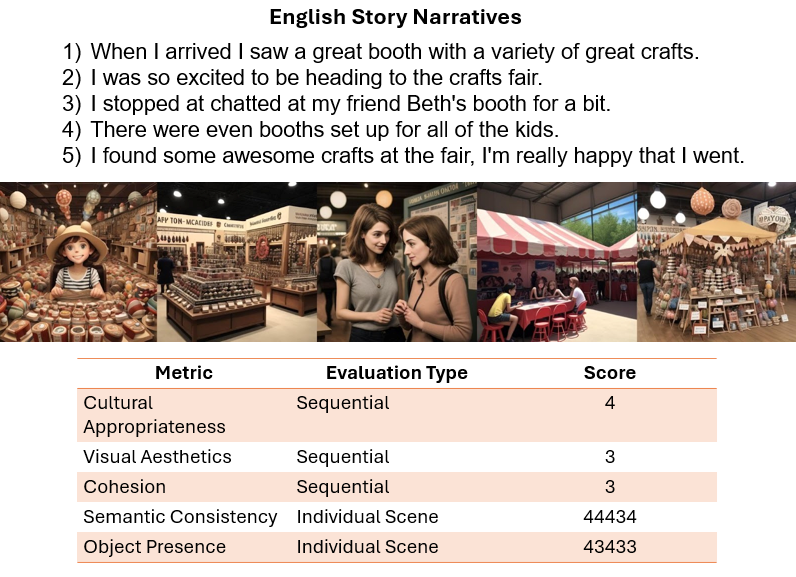}
  \caption{An example of human evaluation results for an English story across five evaluation metrics.}
  \label{fig:ha-english}
\end{figure}

\begin{figure}[ht]
  \centering
  \includegraphics[width=0.48\textwidth]{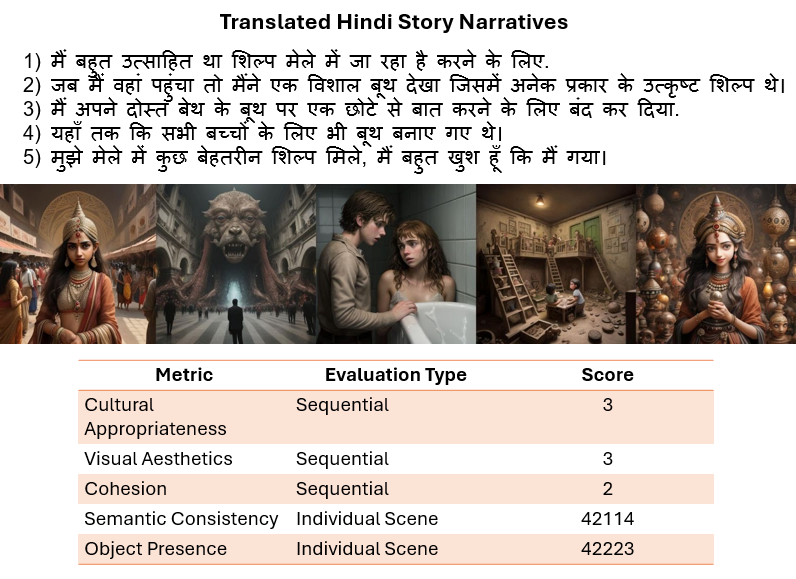}
  \caption{An example of human evaluation results for a Hindi story across five evaluation metrics.}
  \label{fig:ha-hindi}
\end{figure}

\begin{figure}[h]
  \centering
  \includegraphics[width=0.48\textwidth]{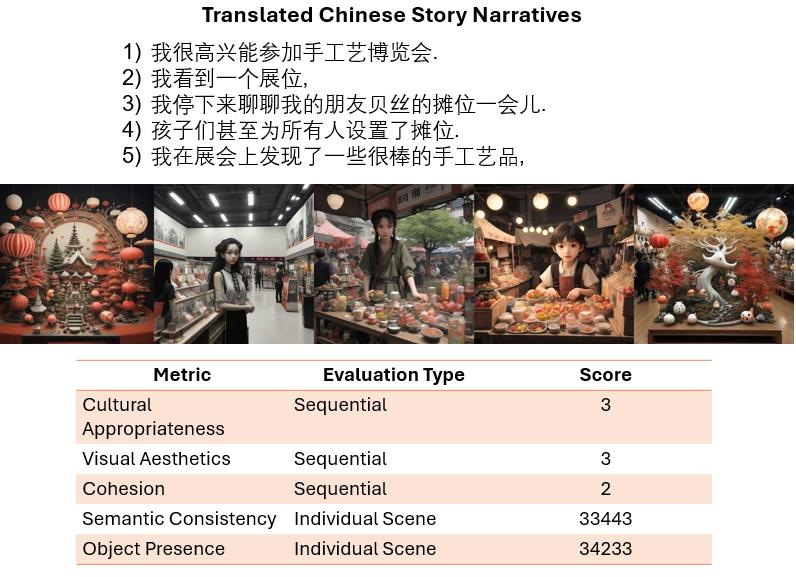}
  \caption{An example of human evaluation results for a Chinese story across five evaluation metrics.}
  \label{fig:ha-Chinese}
\end{figure}

Sequential metrics such as Cultural Appropriateness, Visual Aesthetics and Cohesion were rated with a single score over the entire story, which contributed to lower agreement among annotators due to their subjective nature. In contrast, individual scene metrics like Semantic Consistency and Object Presence were evaluated separately for each scene, leading to higher inter-annotator agreement as this more granular approach reduces ambiguity and improves consistency. Figures~\ref{fig:ha-english}, \ref{fig:ha-hindi}, and \ref{fig:ha-Chinese} show examples of human evaluation results for the same story across three sequential and two individual scene evaluation metrics in English, Hindi, and Chinese, respectively.

\begin{table}[h]
    \centering
    \small
    \setlength{\tabcolsep}{2pt} 
    \renewcommand{\arraystretch}{1.2}
    \begin{tabular}{@{}lccc@{}}
        \toprule
        \textbf{Metric} & \textbf{English ($\kappa$)} & \textbf{Hindi ($\kappa$)} & \textbf{Chinese ($\kappa$)} \\
        \midrule
        Cultural Appropriateness & 0.239 & 0.262 & 0.164 \\
        Visual Aesthetics        & 0.143 & 0.263 & 0.118 \\
        Cohesion                 & 0.257 & 0.231 & 0.280 \\
        \rowcolor{orange!15}
        Semantic Consistency     & 0.551 & 0.610 & 0.422 \\
        \rowcolor{orange!15}
        Object Presence          & 0.525 & 0.605 & 0.340 \\
        \bottomrule
    \end{tabular}
    \caption{Weighted Cohen’s $\kappa$ scores for each evaluation metric across English, Hindi and Chinese.}
    \label{tab:kappa_scores}
\end{table}

Our goal was to capture a broad spectrum of expert opinions rather than enforce consensus, so we do not emphasize inter-rater reliability but instead present aggregated expert ratings across all metrics (see Figure 3 in the main paper). All evaluators were volunteers with relevant academic backgrounds and received a detailed briefing on rating guidelines to promote a consistent understanding of the evaluation metrics across languages.

\renewcommand{\arraystretch}{1.1}

\begin{table}[ht]
\centering
\resizebox{0.48\textwidth}{!}{%
\begin{tabular}{
  l
  *{3}{c}
  *{3}{c}
}
\toprule
\multicolumn{1}{l}{\Large\textbf{Models}} &
\multicolumn{3}{c}{\Large\textbf{CLIP $\uparrow$}} &
\multicolumn{3}{c}{\Large\textbf{FID $\downarrow$}} \\
\cmidrule(lr){2-4} \cmidrule(lr){5-7}

& \large\textbf{English} & \large\textbf{Hindi} & \large\textbf{Chinese}
& \large\textbf{English} & \large\textbf{Hindi} & \large\textbf{Chinese} \\

\rowcolor{gray!25}
\multicolumn{7}{c}{\large\textbf{FlintstonesSV}} \\

\large MuLan SD2v1     
  & 29.18 & 25.73 & 27.54
  & 216.23 & 230.63 & 236.97 \\

\large MuLan SD1v5     
  & 28.71 & 26.32 & 27.18
  & 251.07 & 252.19 & 259.31   \\

\rowcolor{orange!15}
\large MuLan SDXL     
   & 27.94 & 25.50 & 27.31
   & 239.52 & 242.31 & 250.70 \\

\rowcolor{blue!15}
\large Average     
   & 28.61 & 25.85 & 27.34
   & 235.61 & 242.31 & 250.70 \\

\addlinespace
\rowcolor{gray!25}
\multicolumn{7}{c}{\large\textbf{VIST}} \\

\large MuLan SD2v1     
    & 25.62 & 23.95 & 24.22
   & 86.21 & 94.11 & 103.95 \\

\large MuLan SD1v5     
   & 25.77 & 24.05 & 24.03 
   & 87.42 & 97.41 & 113.27 \\

\rowcolor{orange!15}
\large MuLan SDXL      
  & 25.82 & 23.88 & 24.51
   & 89.23 & 93.61 & 96.82 \\

\rowcolor{blue!15}
\large Average     
   & 25.74 & 23.96 & 24.25
   & 87.62 & 95.04 & 104.68 \\

\bottomrule
\end{tabular}
}
\caption{Results of CLIP and FID score on story generated on VIST and FlintstonesSV datasets}
\label{tab:CLIP_and_FID_results}
\end{table}

% \subsection{Culture Error Analysis}
% \label{sec:culture_error_analysis}
% Figure~\ref{fig:Hindi_Error_analysis_Sample} shows an example from a Hindi story depicting a family vacation with friends. In Scene 1, the absence of appropriate footwear reflects a stereotypical depiction of Indian culture, while Scene 4 exhibits an attire misclassification error, where the generated clothing does not correspond to traditional Indian styles. Similarly, Figure~\ref{fig:English_Error_analysis_Sample} presents an example from a Chinese story about families enjoying a game together. In Scene 2, the flag resembles an American flag instead of a Chinese one, representing a Symbol Mislocalization error. In Scene 4, the exaggerated height of the male character compared to the woman and girl reflects a Social Role Misrepresentation. Finally, Scene 5 shows multiple characters with overly narrowed eyes and nearly identical expressions, indicating culturally inappropriate stereotype amplification.

% whilke family gathered to watch the gamea group of fans visiting a baseball stadium to watch their favorite team play. In Scene 3, the appearance of a dog instead of a player on the baseball field represents a Symbol Mislocalization error, highlighting the model’s confusion in associating the correct object with the scene. Scene 4 shows the baseball player throwing the bat instead of using it to hit the ball, which represents a Social Role Misinterpretation error, indicating the model’s difficulty in associating the correct roles within the scene.

\begin{figure}[htpb]
  \centering
  \includegraphics[width=0.48\textwidth]{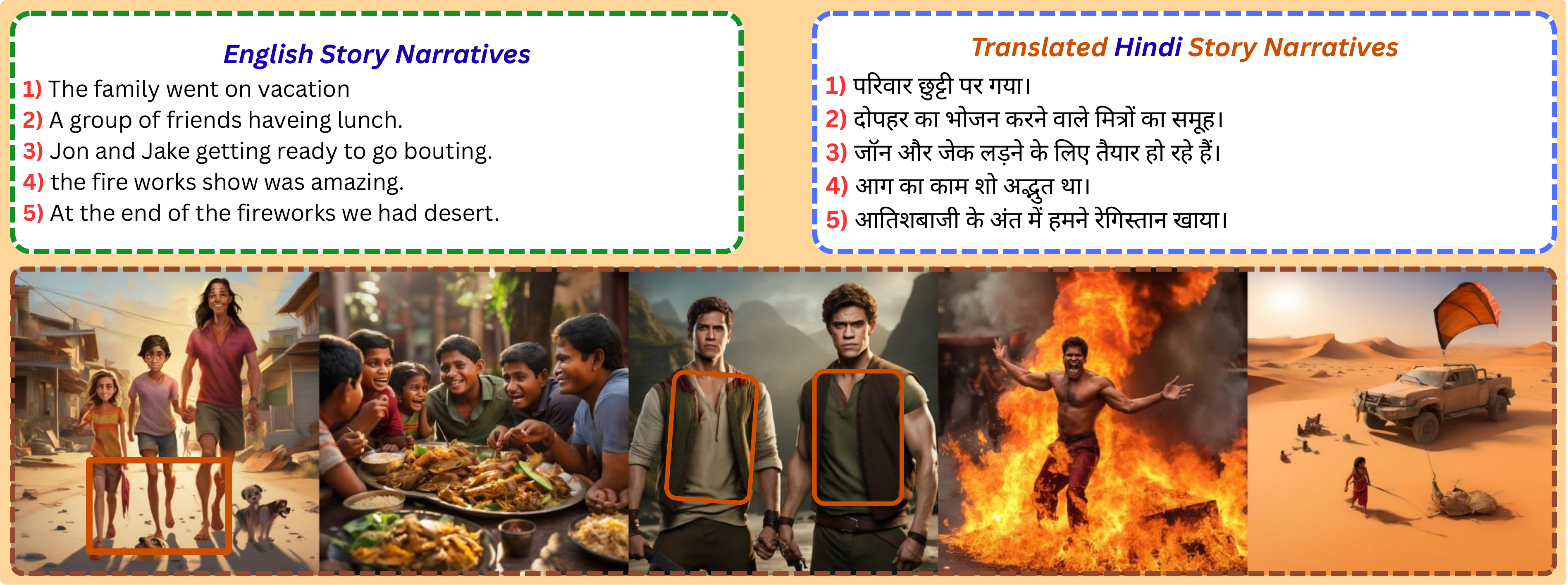}
  \caption{Hindi Error Analysis on the Visualization of the same story in different languages generated by \texttt{MuLan-SDXL} model depicted with orange boxes.}
  \label{fig:Hindi_Error_analysis_Sample}
\end{figure}

\begin{figure}[htpb]
  \centering
  \includegraphics[width=0.48\textwidth]{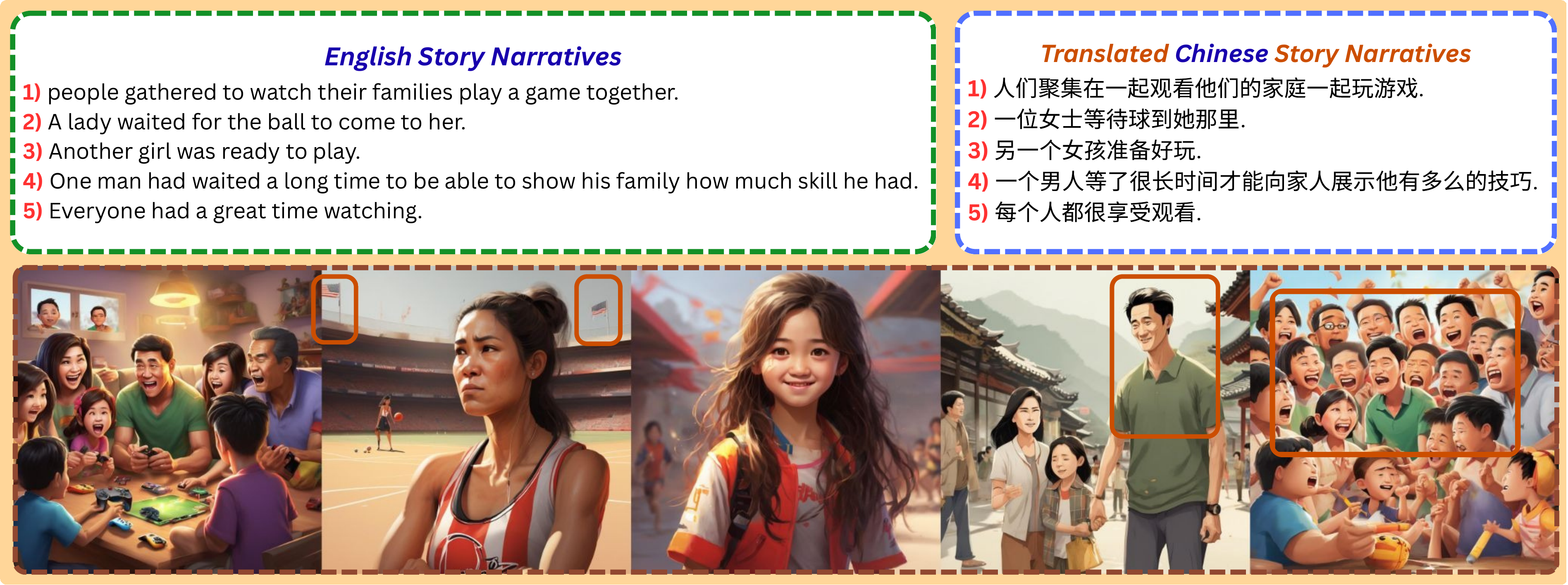}
  \caption{Chinese Error Analysis on the Visualization of the same story in different languages generated by \texttt{MuLan-SDXL} model depicted with orange boxes.}
  \label{fig:English_Error_analysis_Sample}
\end{figure}

% ------------------------------

\begin{figure*}[htpb]
  \centering
  \includegraphics[width=\textwidth]{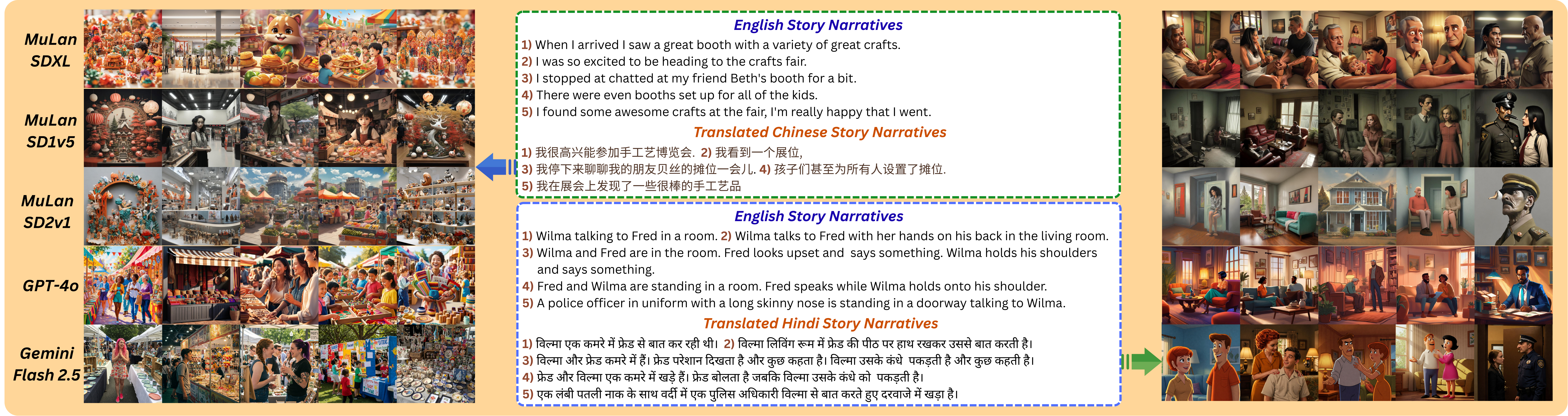}
  \caption{Visualization of the same story generated by different models in the same language.  \textbf{Left}: Chinese  \textbf{Right}: Hindi}
  \label{fig:same_story_across_different_models}
\end{figure*}

\begin{figure*}[ht]
  \centering
  \includegraphics[width=0.98\textwidth]{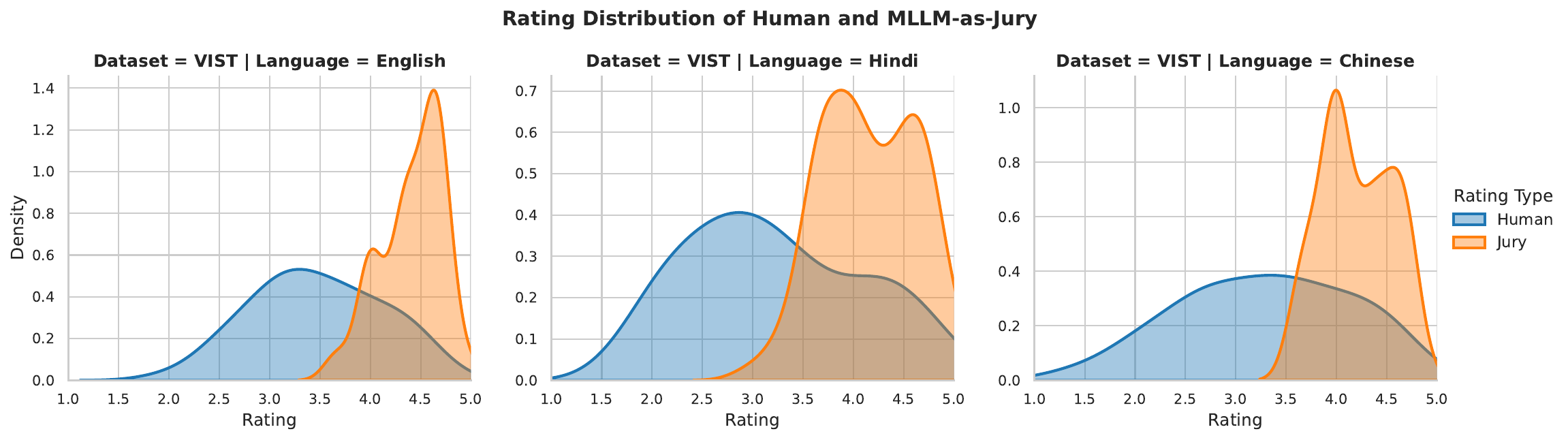}
  \hfill
  \includegraphics[width=0.98\textwidth]{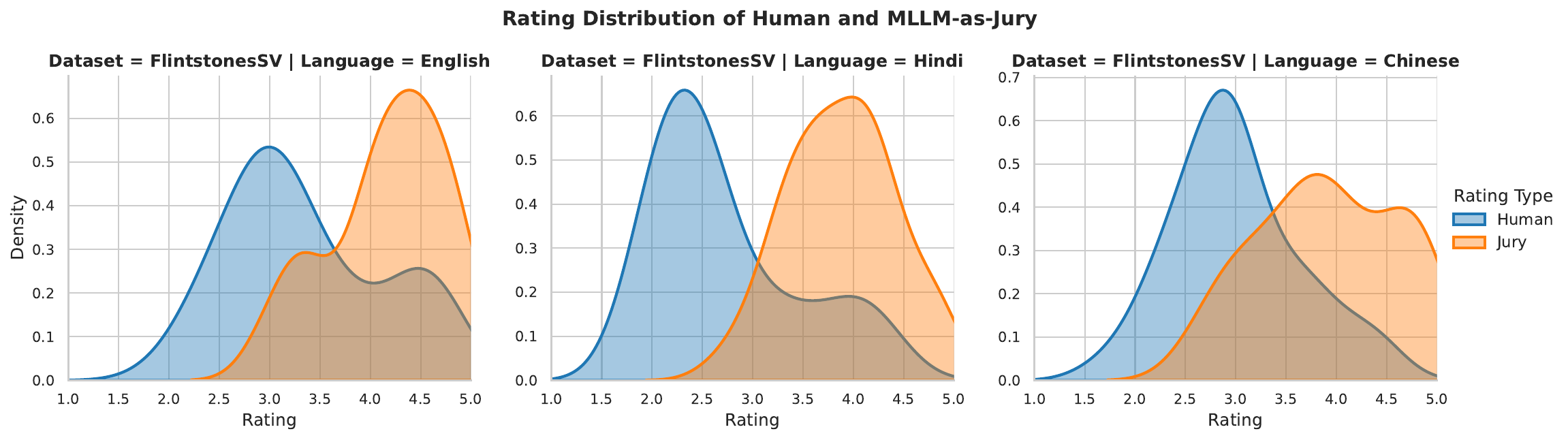}
  \caption{Rating distribution of Human and MLLM-as-Jury method across dataset and languages.}
  \label{fig:Rating_Distribution_Plots}
\end{figure*}

\begin{figure*}[ht]
  \centering
  \includegraphics[width=\textwidth]{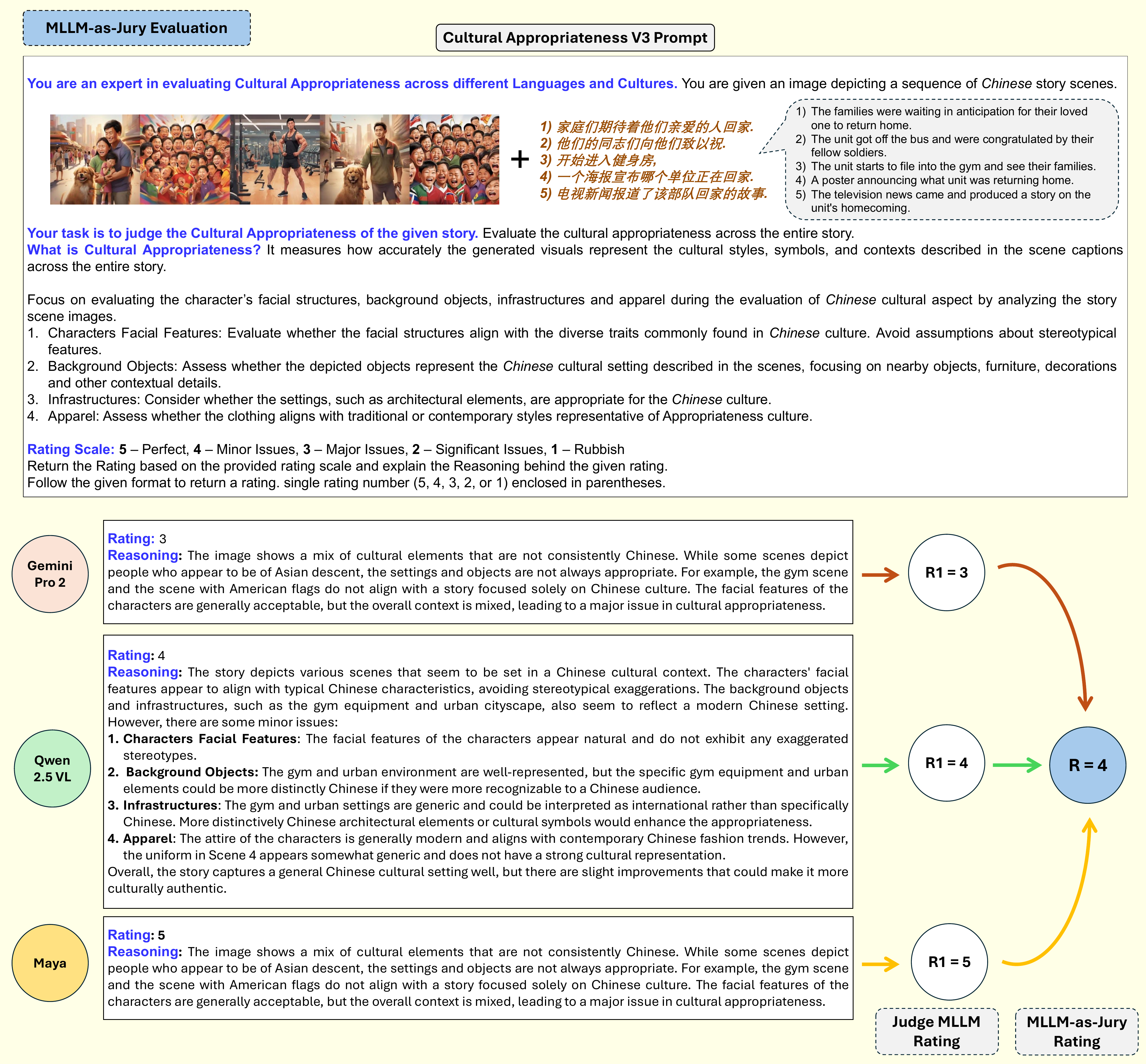}
  \caption{Schematic representation of the MLLM-as-Jury evaluation on Cultural Appropriateness V3 in Chinese using three MLLMs: Gemini Pro 2, Qwen 2.5 VL, and Maya.}
  \label{fig:MLLM-as-Jury_workflow}
\end{figure*}

\begin{figure*}[ht]
  \centering
  \includegraphics[width=\textwidth]{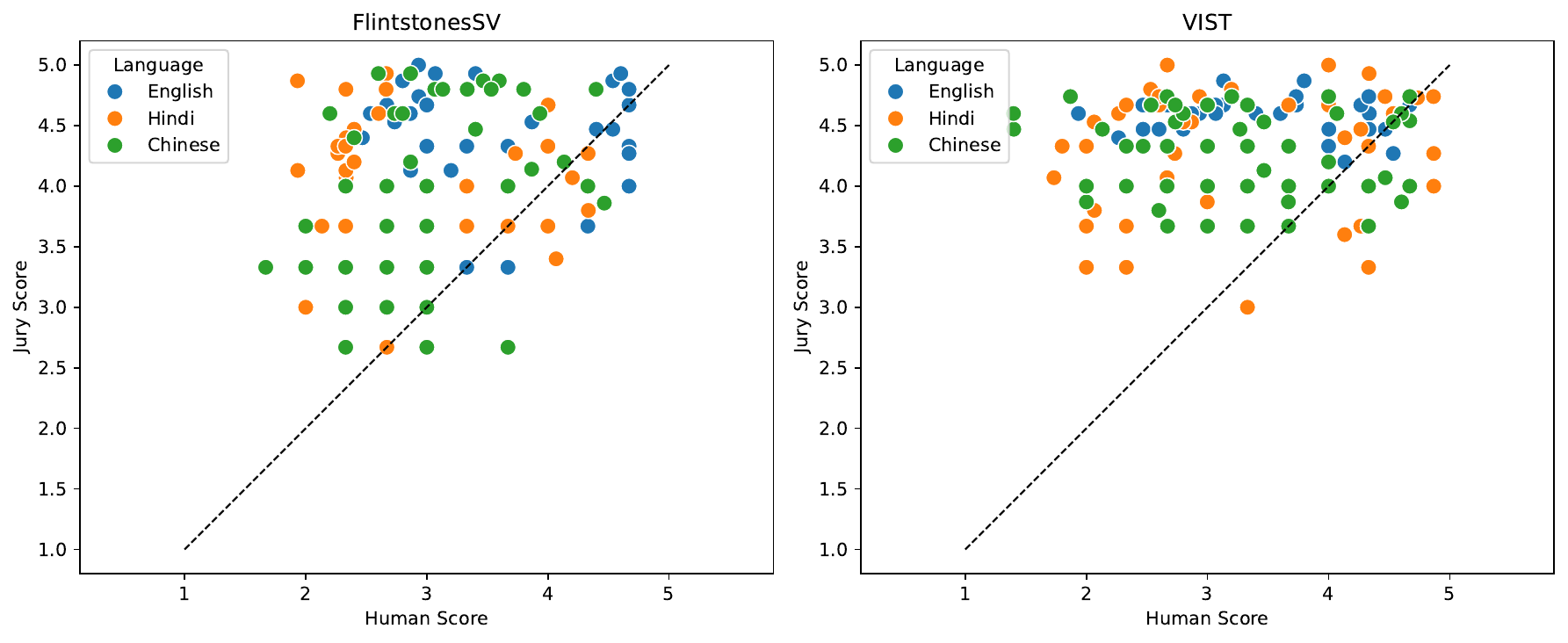}
  \caption{Scatter plot of ratings of Human and MLLM-as-Jury evalaution on FlintstonesSV and VIST datasets. }
  \label{fig:Human_vs_Jury_Scatter_Plot}
\end{figure*}

\subsection{MLLM-as-Jury Ablation Study}
\label{MLLM-as-Jury Ablation Study}
\begin{figure*}[ht]
  \centering
  \includegraphics[width=\textwidth]{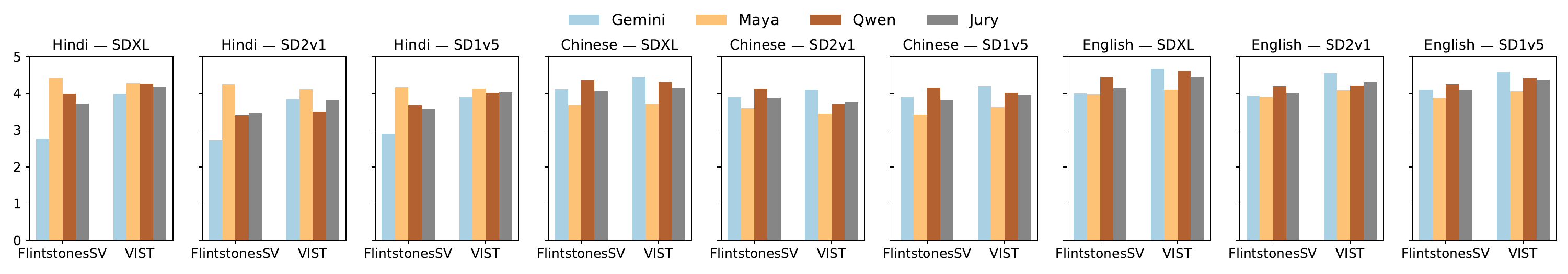}
  \caption{Semantic Consistency Jury vs Judge Results}
  \label{fig:Judge_vs_Jury_Semantic_Consistency}
\end{figure*}

\begin{figure*}[ht]
  \centering
  \includegraphics[width=\textwidth]{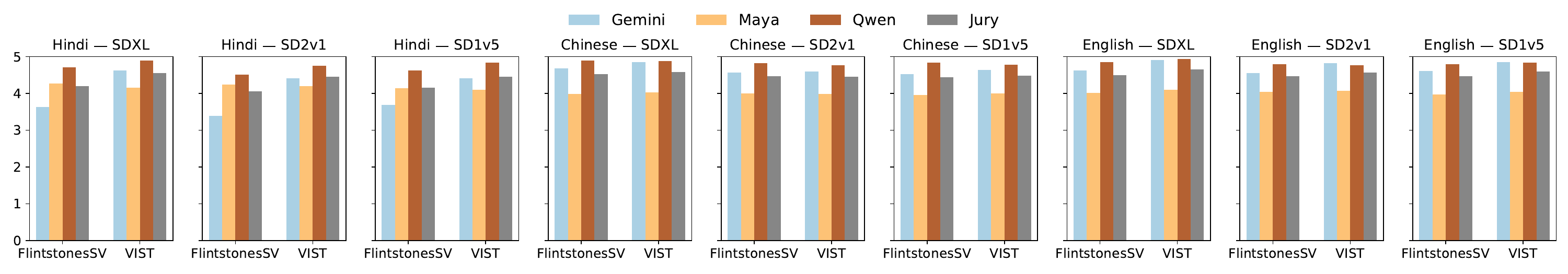}
  \caption{Object Presence Jury vs Judge Results}
  \label{fig:Judge_vs_Jury_Object_Presence}
\end{figure*}

\begin{figure*}[ht]
  \centering
  \includegraphics[width=\textwidth]{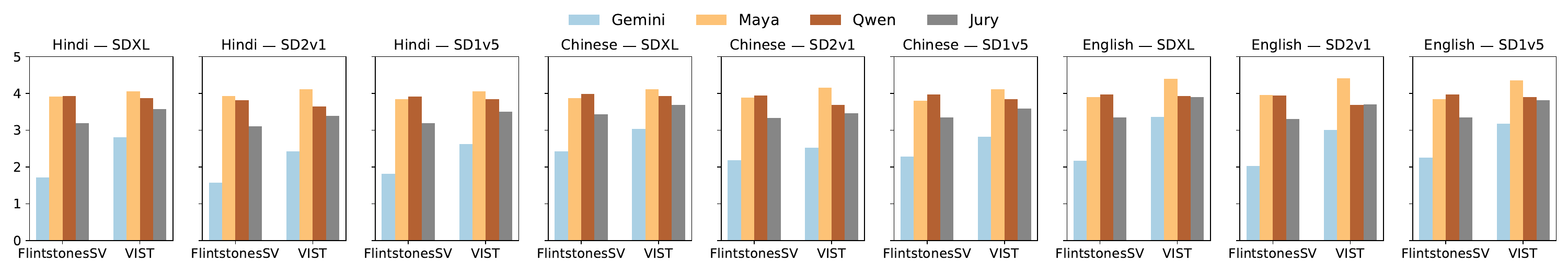}
  \caption{Cohesion Jury vs Judge Results}
  \label{fig:Judge_vs_Jury_Cohesion}
\end{figure*}

\begin{figure*}[ht]
  \centering
  \includegraphics[width=\textwidth]{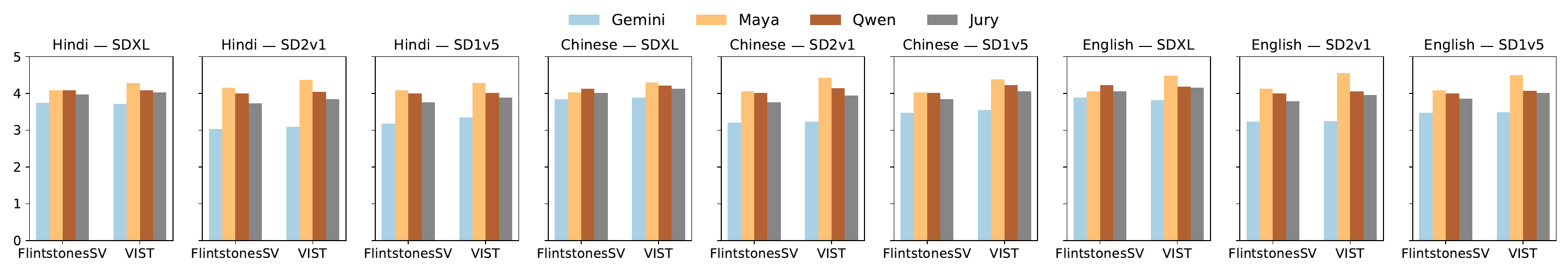}
  \caption{Visual Aesthetics Jury vs Judge Results}
  \label{fig:Judge_vs_Jury_Visual_Aesthetics}
\end{figure*}

This section compares the single-judge models with the jury method to examine how individual judges contribute across languages, metrics, and datasets. For Semantic Consistency in Figure \ref{fig:Judge_vs_Jury_Semantic_Consistency} and Object Presence in Figure \ref{fig:Judge_vs_Jury_Object_Presence} show that single-judge scores and jury scores are largely aligned across all languages. For Cohesion, Figure \ref{fig:Judge_vs_Jury_Cohesion} shows that Gemini consistently assigns lower scores than Maya and Qwen across all languages, particularly on FlintstonesSV compared to VIST. The jury method mitigates this bias by averaging the judges’ outputs. For Visual Aesthetics, Gemini again scores moderately lower than Qwen and Maya, though the gap is smaller than in the Cohesion metric, as seen in Figure \ref{fig:Judge_vs_Jury_Visual_Aesthetics}.

\begin{figure*}[ht]
  \centering
  \includegraphics[width=\textwidth]{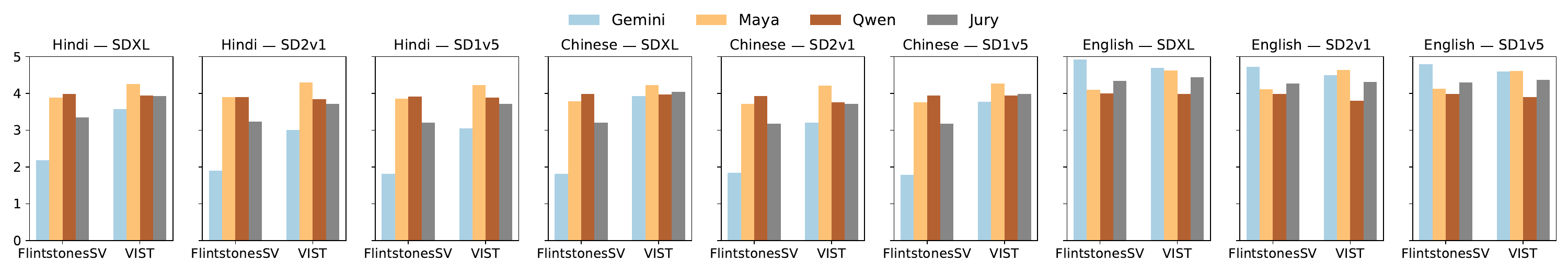}
  \caption{Cultural Appropriateness V1 - Jury vs Judge Results}
  \label{fig:Judge_vs_Jury_Cultural_Appropriateness_V1}
\end{figure*}

\begin{figure*}[ht]
  \centering
  \includegraphics[width=\textwidth]{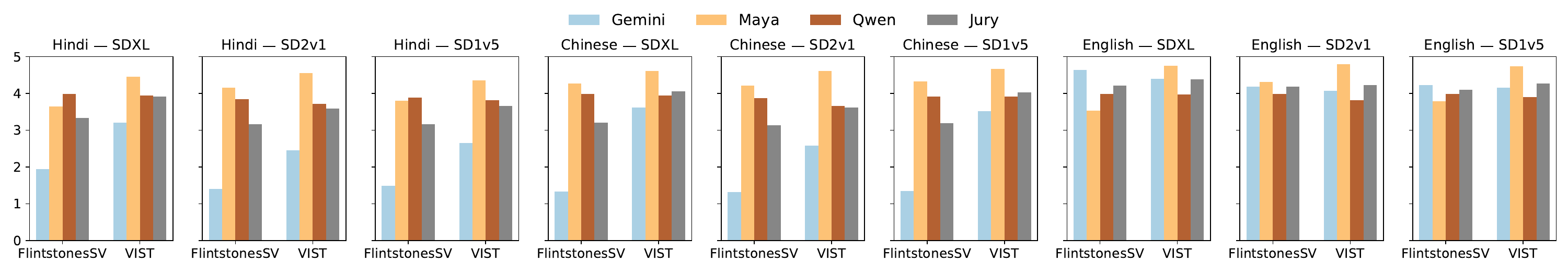}
  \caption{Cultural Appropriateness V2 - Jury vs Judge Results}
  \label{fig:Judge_vs_Jury_Cultural_Appropriateness_V2}
\end{figure*}

\begin{figure*}[ht]
  \centering
  \includegraphics[width=\textwidth]{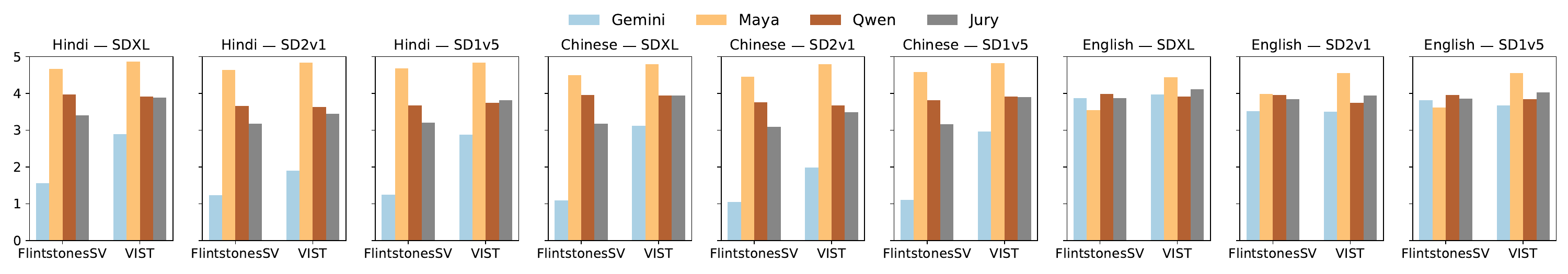}
  \caption{Cultural Appropriateness V3 - Jury vs Judge Results}
  \label{fig:Judge_vs_Jury_Cultural_Appropriateness_V3}
\end{figure*}

A different pattern emerges for Cultural Appropriateness (Figures \ref{fig:Judge_vs_Jury_Cultural_Appropriateness_V1}, \ref{fig:Judge_vs_Jury_Cultural_Appropriateness_V2}, and \ref{fig:Judge_vs_Jury_Cultural_Appropriateness_V3}). Gemini assigns the lowest scores on Hindi and Chinese, indicating weak cultural fidelity in these languages in generated stories, and FlintstonesSV receives consistently lower scores than VIST, showing real-world story adapts to culture more than animated ones. As prompt complexity increases from V1 to V2 and from V2 to V3, the scores of Gemini and Qwen decrease, indicating that the added cultural focus points and illustrative examples enforce a stricter evaluation of cultural aspects. In contrast, Maya’s scores remain largely stable. For English, Gemini’s scores are higher than for Hindi and Chinese. Overall, Maya assigns substantially higher scores than the other two judges at V2 and V3, especially on VIST. The jury method reduces this imbalance by averaging the three judges’ scores, though some outlier influence persists on VIST, as reflected in the rating distribution shown in Figure \ref{fig:Rating_Distribution_Plots}. Overall, across all metrics, the MLLM-as-Jury approach provides a more stable and reliable automatic evaluation for rubric-based story visualization by reducing individual model bias and leveraging complementary judging capabilities. Figure~\ref{fig:MLLM-as-Jury_workflow} shows a schematic representation of the MLLM-as-Jury evaluation on Cultural Appropriateness V3 in Chinese using the three MLLMs.

\subsection{Prompts used in MLLM-as-Jury Evaluation}

This section presents the prompts used in our MLLM-as-Jury evaluation framework. Figure~\ref{fig:prompt_semantic} shows the prompt for assessing semantic consistency, which ensures logical alignment between visual and textual content. Figure~\ref{fig:prompt_object} contains the prompt for evaluating object presence, verifying whether the described entities appear in the image. Visual aesthetics are evaluated using the prompt in Figure~\ref{fig:prompt_aesthetics}, which focuses on image quality and composition. Cohesion is assessed through the prompt in Figure~\ref{fig:prompt_cohesion}, which measures narrative flow and continuity.

Our Progressive Multicultural Evaluation framework enables culturally grounded assessment through three prompt versions. Version 1 (V1) (Figure~\ref{fig:prompt_cultural_v1}) introduces a culture definition that establishes the criteria for cultural representation. Version 2 (V2) (Figure~\ref{fig:prompt_cultural_v2}) extends this with focused cultural aspects. Version 3 (V3) (Figure~\ref{fig:prompt_cultural_v3}) further refines the evaluation by adding illustrative examples, providing concrete references that help MLLMs produce more accurate and consistent cultural judgments.
%Using diverse models in the evaluation process makes the framework more robust, as it combines different strengths and perspectives in assessing story quality.

\subsection{MLLM-as-Jury Models Configuration}
In this section, we describe the configuration of our MLLM-as-Jury evaluation framework, designed to enable large-scale, automated assessment of various story aspects. As introduced in the main paper, this framework leverages three Multimodal Large Language Models (MLLMs): \texttt{Gemini Pro 2.0}, \texttt{Qwen2.5-VL} and \texttt{Maya}. Table~\ref{tab:vlm_params} presents the configuration details and parameters used for each model. Using diverse models in the evaluation process makes the framework more robust, as it combines different strengths and perspectives in assessing story quality.

\begin{table}[htbp]
    \centering
    \small
    \renewcommand{\arraystretch}{1.2}
    \begin{tabularx}{0.45\textwidth}{@{}l X X X@{}}
        \toprule
        \textbf{Parameter} & \textbf{Gemini Pro 2.0} & \textbf{QwenVL 2.5} & \textbf{Maya} \\
        \midrule
        Access & Gemini API & GPU Inference & GPU Inference \\
        Max Tokens & 512 & 512 & 512 \\
        Sampling & False & False & False \\
        Temperature & 0 & 0 & 0 \\
        Prompt & Custom & Custom & Custom \\
        Seed & 42 & 42 & 42 \\
        Hardware & RTX A6000 48GB GPU & RTX A6000 48GB GPU & RTX A6000 48GB GPU \\
        \bottomrule
    \end{tabularx}
    \caption{Inference configuration for Gemini Pro 2.0, QwenVL 2.5 and Maya models.}
    \label{tab:vlm_params}
\end{table}

\begin{table}[h]
    \centering
    \small
    \setlength{\tabcolsep}{2.5pt} 
    \renewcommand{\arraystretch}{1.2}
    \begin{tabular}{@{}p{0.28\linewidth} p{0.22\linewidth} p{0.22\linewidth} p{0.22\linewidth}@{}}
        \toprule
        \textbf{Parameter} & \textbf{SDXL} & \textbf{SD2v1} & \textbf{SD1v5} \\
        \midrule
        Multilingual Adapter & MuLan & MuLan & MuLan \\
        Guidance Scale & 7 & 7 & 7 \\
        Inference Steps & 50 & 50 & 50 \\
        Framework Used & Diffusers & Diffusers & Diffusers \\
        Seed  & 12345 & 12345 & 12345 \\
        Hardware & RTX A6000 48GB GPU & RTX A6000 48GB GPU & RTX A6000 48GB GPU \\
        \bottomrule
    \end{tabular}
    \caption{Inference configuration for Text-to-Image models used: MuLan-SDXL, MuLan-SD2v1 and MuLan-SD1v5}
    \label{tab:t2i_params}
\end{table}

\subsection{Human vs MLLM-as-Jury Correlation}
Figure~\ref{fig:Human_vs_Jury_Scatter_Plot} shows scatter plots comparing human ratings with MLLM-as-Jury ratings for the FlintstonesSV and VIST datasets across three languages. For FlintstonesSV, the spread of ratings lies moderately along the diagonal, indicating a moderate correlation with human evaluations, as also shown in Figure~\ref{fig:Human_vs_Jury_Correlation_Scores}. In contrast, for VIST, the rating distribution is more horizontally spread on the higher side of the MLLM-as-Jury scores, with no clear diagonal trend, resulting in very low correlation. This suggests that MLLM-as-Jury is less critical for open-ended, contextually rich, and culturally nuanced scenes, often assigning higher scores. However, for simpler tasks with fixed settings, such as FlintstonesSV, it evaluates more critically and aligns more closely with human judgments. Overall, the results show that MLLM-as-Jury performance is dataset-dependent and influenced by the choice of underlying judge models.

%Figure~\ref{fig:Human_vs_Jury_Scatter_Plot} shows scatter plots comparing human ratings with MLLM-as-Jury ratings for the FlintstonesSV and VIST datasets across three languages. For FlintstonesSV, the spread of ratings lies moderately along the diagonal, reflecting a moderate correlation between MLLM-as-Jury and human evaluations, as shown in Figure~\ref{fig:Human_vs_Jury_Correlation_Scores}. In contrast, for VIST, the rating distribution is more horizontally spread on the higher side of the MLLM-as-Jury ratings, with no clear diagonal trend, leading to a very low correlation with human judgments. This analysis indicates that MLLM-as-Jury tends to be less critical when evaluating open-ended, contextually rich, and culturally nuanced scenes, often assigning higher scores. However, for simpler tasks with fixed settings, such as in FlintstonesSV, it evaluates more critically and produces ratings closer to human evaluations. Overall, the results demonstrate that MLLM-as-Jury performance is dataset-dependent and influenced by the choice of the underlying MLLM-as-Judge models.

\subsection{Multilingual Text-to-Image Models Configuration}
This section provides the configuration details for the multilingual text-to-image (T2I) models used to generate story from scene-level narratives in various languages. All models are based on the MuLan framework, which uses language adapters to extend English-centric diffusion models for multilingual prompts. Three variants of MuLan-enabled Stable Diffusion were used in our experiments: SD1v5, SD2v1 and SDXL. These models differ in architecture and image resolution, enabling a range of visual generation capabilities. Table~\ref{tab:t2i_params} outlines the inference configurations for each model, including resolution, scheduler type and hardware settings. The remaining two models, GPT-4o and Gemini-Flash-2.0, were used with their default settings through the chat interface.

% =====================================
\begin{figure*}[!ht]
\noindent
\begin{tcolorbox}[
  colback=white,
  colframe=black,
  arc=2mm,
  boxrule=0.5pt,
  width=\textwidth,
  toptitle=1pt,
  bottomtitle=1pt,
  title=\bfseries Semantic Consistency Prompt
]
\ttfamily
You are an expert in evaluating \textbf{Semantic Consistency in Visual Stories.}\\

You are given an image depicting a sequence of \textbf{\{language}\} story scenes.\\

\textbf{\{language\} Story Scene Descriptions:}\\

Scene 1: \{scene\_descriptions[0]\} \\
Scene 2: \{scene\_descriptions[1]\} \\
Scene 3: \{scene\_descriptions[2]\} \\
Scene 4: \{scene\_descriptions[3]\} \\
Scene 5: \{scene\_descriptions[4]\} \\

Your task is to carefully check if \textbf{scene description is semantically consistent with scene image or not}.
 \\

\textbf{Rating Scale:}

5 – Perfect \\
4 – Minor Issues \\
3 – Major Issues \\
2 – Significant Issues \\
1 – Rubbish\\

Return the Rating based on the provided rating scale and explain the Reasoning behind the given rating.\\

If any scene description is not semantically meaningful to scene image, lower the rating accordingly. Do not give a perfect score unless its fully semantically consistent. \\

Strictly first return the rating in the format \textbf{(X, X, X, X, X)}, where each X is a single rating number \textbf{(5, 4, 3, 2, or 1)} corresponding to each story scene followed by reasoning.
\\

\textbf{Reasoning:}
\end{tcolorbox}
\caption{Prompt used for evaluating Semantic Consistency in MLLM-as-Jury Evaluation}
\label{fig:prompt_semantic}
\end{figure*}

% =====================================
\begin{figure*}[!ht]
\noindent
\begin{tcolorbox}[
  colback=white,
  colframe=black,
  arc=2mm,
  boxrule=0.5pt,
  width=\textwidth,
  toptitle=1pt,
  bottomtitle=1pt,
  title=\bfseries Object Presence Prompt
]
\ttfamily
You are an expert in evaluating \textbf{Object Presence in Visual Stories.}\\

You are given an image depicting a sequence of \textbf{\{language}\} story scenes.\\

\textbf{\{language\} Story Scene Descriptions:}\\

Scene 1: \{scene\_descriptions[0]\} \\
Scene 2: \{scene\_descriptions[1]\} \\
Scene 3: \{scene\_descriptions[2]\} \\
Scene 4: \{scene\_descriptions[3]\} \\
Scene 5: \{scene\_descriptions[4]\} \\

Your task is to carefully check if every \textbf{object mentioned in the scene description is clearly present and visible} in the corresponding scene image.
 \\

\textbf{Rating Scale:}

5 – Perfect \\
4 – Minor Issues \\
3 – Major Issues \\
2 – Significant Issues \\
1 – Rubbish\\

Return the Rating based on the provided rating scale and explain the Reasoning behind the given rating.\\

If any object is missing or unclear, lower the rating accordingly. Do not give a perfect score unless all objects are fully and clearly present.\\

Strictly first return the rating in the format \textbf{(X, X, X, X, X)}, where each X is a single rating number \textbf{(5, 4, 3, 2, or 1)} corresponding to each story scene followed by reasoning.
\\

\textbf{Reasoning:}
\end{tcolorbox}
\caption{Prompt used for evaluating Object Presence in MLLM-as-Jury Evaluation}
\label{fig:prompt_object}
\end{figure*}

% =====================================

% =====================================
\begin{figure*}[!ht]
\noindent
\begin{tcolorbox}[
  colback=white,
  colframe=black,
  arc=2mm,
  boxrule=0.5pt,
  width=\textwidth,
  toptitle=1pt,
  bottomtitle=1pt,
  title=\bfseries Visual Aesthetics Prompt
]
\ttfamily
You are an expert in evaluating \textbf{Visual Aesthetics in Visual Stories.}.\\

You are given an image depicting a sequence of \textbf{\{language\}} story scenes. \\

\textbf{\{language\} Story Scene Descriptions:}\\

Scene 1: \{scene\_descriptions[0]\} \\
Scene 2: \{scene\_descriptions[1]\} \\
Scene 3: \{scene\_descriptions[2]\} \\
Scene 4: \{scene\_descriptions[3]\} \\
Scene 5: \{scene\_descriptions[4]\} \\

Your task is to judge the \textbf{Visual Aesthetics} of the given story. Evaluate the Visual Aesthetics across the entire story.
 \\

\textbf{What is Visual Aesthetics?} It measures the coherence and logical flow between scenes, ensuring smooth transitions and meaningful relationships in the visual narrative.\\

\textbf{Rating Scale:}

5 – Perfect \\
4 – Minor Issues \\
3 – Major Issues \\
2 – Significant Issues \\
1 – Rubbish\\

Return the Rating based on the provided rating scale and explain the Reasoning behind the given rating.\\

\textbf{Follow the below given format to return Rating. }\\

single rating number (5, 4, 3, 2, or 1) enclosed in parentheses. \\

\textbf{Reasoning:}
\end{tcolorbox}
\caption{Prompt used for evaluating Visual Aesthetics in MLLM-as-Jury Evaluation}
\label{fig:prompt_aesthetics}
\end{figure*}

% =====================================
\begin{figure*}[!ht]
\noindent
\begin{tcolorbox}[
  colback=white,
  colframe=black,
  arc=2mm,
  boxrule=0.5pt,
  width=\textwidth,
  toptitle=1pt,
  bottomtitle=1pt,
  title=\bfseries Cohesion Prompt
]
\ttfamily
You are an expert in evaluating \textbf{Cohesion in Visual Stories}.\\

You are given an image depicting a sequence of \textbf{\{language\}} story scenes. \\

\textbf{\{language\} Story Scene Descriptions:}\\

Scene 1: \{scene\_descriptions[0]\} \\
Scene 2: \{scene\_descriptions[1]\} \\
Scene 3: \{scene\_descriptions[2]\} \\
Scene 4: \{scene\_descriptions[3]\} \\
Scene 5: \{scene\_descriptions[4]\} \\

Your task is to evaluate the \textbf{Cohesion} of the given story across all scenes. \\

\textbf{What is Cohesion?} It measures the coherence and logical flow between scenes, ensuring smooth transitions and meaningful relationships in the visual narrative.\\

\textbf{Rating Scale:}

5 – Perfect \\
4 – Minor Issues \\
3 – Major Issues \\
2 – Significant Issues \\
1 – Rubbish\\

Return the Rating based on the provided rating scale and explain the Reasoning behind the given rating.\\

\textbf{Follow the below given format to return Rating. }\\

single rating number (5, 4, 3, 2, or 1) enclosed in parentheses. \\

\textbf{Reasoning:}
\end{tcolorbox}
\caption{Prompt used for evaluating Cohesion in MLLM-as-Jury Evaluation}
\label{fig:prompt_cohesion}
\end{figure*}

% =====================================

% =====================================
\begin{figure*}[!ht]
\noindent
\begin{tcolorbox}[
  colback=white,
  colframe=black,
  arc=2mm,
  boxrule=0.5pt,
  width=\textwidth,
  toptitle=1pt,
  bottomtitle=1pt,
  title=\bfseries Cultural Appropriateness V1 Prompt
]
\ttfamily
You are an expert in evaluating \textbf{Cultural Appropriateness across different Languages and Cultures}.\\

You are given an image depicting a sequence of \textbf{\{language\}} story scenes. \\

\textbf{\{language\} Story Scene Descriptions:}\\

Scene 1: \{scene\_descriptions[0]\} \\
Scene 2: \{scene\_descriptions[1]\} \\
Scene 3: \{scene\_descriptions[2]\} \\
Scene 4: \{scene\_descriptions[3]\} \\
Scene 5: \{scene\_descriptions[4]\} \\

Your task is to judge the \textbf{Cultural Appropriateness} of the given story. Evaluate the Cultural Appropriateness across the entire story. \\

\textbf{What is Cultural Appropriateness?} It measures how accurately the generated visuals represent the cultural styles, symbols and contexts described in the scene captions across the entire story.\\

\textbf{Rating Scale:}

5 – Perfect \\
4 – Minor Issues \\
3 – Major Issues \\
2 – Significant Issues \\
1 – Rubbish\\

Return the Rating based on the provided rating scale and explain the Reasoning behind the given rating.\\

\textbf{Follow the below given format to return Rating. }\\

single rating number (5, 4, 3, 2, or 1) enclosed in parentheses. \\

\textbf{Reasoning:}
\end{tcolorbox}
\caption{Prompt used for evaluating Cultural Appropriateness level V1 in MLLM-as-Jury Evaluation}
\label{fig:prompt_cultural_v1}
\end{figure*}

% =====================================
\begin{figure*}[!ht]
\noindent
\begin{tcolorbox}[
  colback=white,
  colframe=black,
  arc=2mm,
  boxrule=0.5pt,
  width=\textwidth,
  toptitle=1pt,
  bottomtitle=1pt,
  title=\bfseries Cultural Appropriateness V2 Prompt
]
\ttfamily
You are an expert in evaluating \textbf{Cultural Appropriateness across different Languages and Cultures}.\\

You are given an image depicting a sequence of \textbf{\{language\}} story scenes. \\

\textbf{\{language\} Story Scene Descriptions:}\\

Scene 1: \{scene\_descriptions[0]\} \\
Scene 2: \{scene\_descriptions[1]\} \\
Scene 3: \{scene\_descriptions[2]\} \\
Scene 4: \{scene\_descriptions[3]\} \\
Scene 5: \{scene\_descriptions[4]\} \\

Your task is to judge the \textbf{Cultural Appropriateness} of the given story. Evaluate the Cultural Appropriateness across the entire story. \\

\textbf{What is Cultural Appropriateness?} It measures how accurately the generated visuals represent the cultural styles, symbols and contexts described in the scene captions across the entire story.\\

\textbf{Focus} on evaluating the character's facial structures, background objects, infrastructures and apparel during the evaluation of \textbf{\{target\}} cultural Aspect by analyzing the story scene images.\\

\textbf{Rating Scale:}

5 – Perfect \\
4 – Minor Issues \\
3 – Major Issues \\
2 – Significant Issues \\
1 – Rubbish\\

Return the Rating based on the provided rating scale and explain the Reasoning behind the given rating.\\

\textbf{Follow the below given format to return Rating. }\\

single rating number (5, 4, 3, 2, or 1) enclosed in parentheses. \\

\textbf{Reasoning:}
\end{tcolorbox}
\caption{Prompt used for evaluating Cultural Appropriateness level V2 in MLLM-as-Jury Evaluation}
\label{fig:prompt_cultural_v2}
\end{figure*}

% =====================================
\begin{figure*}[!ht]
\noindent
\begin{tcolorbox}[
  colback=white,
  colframe=black,
  arc=2mm,
  boxrule=0.5pt,
  width=\textwidth,
  toptitle=1pt,
  bottomtitle=1pt,
  title=\bfseries Cultural Appropriateness V3 Prompt
]
\ttfamily
You are an expert in evaluating \textbf{Cultural Appropriateness across different Languages and Cultures}.\\

You are given an image depicting a sequence of \textbf{\{language\}} story scenes. \\

\textbf{\{language\} Story Scene Descriptions:}\\

Scene 1: \{scene\_descriptions[0]\} \\
Scene 2: \{scene\_descriptions[1]\} \\
Scene 3: \{scene\_descriptions[2]\} \\
Scene 4: \{scene\_descriptions[3]\} \\
Scene 5: \{scene\_descriptions[4]\} \\

Your task is to judge the \textbf{Cultural Appropriateness} of the given story. Evaluate the Cultural Appropriateness across the entire story. \\

\textbf{What is Cultural Appropriateness?} It measures how accurately the generated visuals represent the cultural styles, symbols and contexts described in the scene captions across the entire story.\\

\textbf{Focus} on evaluating the character's facial structures, background objects, infrastructures and apparel during the evaluation of \textbf{\{target\}} cultural aspect by analyzing the story scene images.\\

1. \textbf{Characters' Facial Features}: Evaluate whether the facial structures align with the diverse traits commonly found in \textbf{\{target\}} culture. Avoid assumptions about stereotypical features.  \\
2. \textbf{Background Objects}: Assess whether the depicted objects represent the \textbf{\{target\}}  cultural setting described in the scenes, focusing on nearby objects, furniture, decorations and other contextual details. \\
3. \textbf{Infrastructures}: Consider whether the settings, such as architectural elements, are appropriate for the \textbf{\{target\}} culture.  \\
4. \textbf{Apparel}: Assess whether the clothing aligns with traditional or contemporary styles representative of \textbf{Appropriate} culture.\\

\textbf{Rating Scale:}

5 – Perfect \\
4 – Minor Issues \\
3 – Major Issues \\
2 – Significant Issues \\
1 – Rubbish\\

Return the Rating based on the provided rating scale and explain the Reasoning behind the given rating.\\

\textbf{Follow the below given format to return Rating. }\\
single rating number (5, 4, 3, 2, or 1) enclosed in parentheses. \\
\textbf{Reasoning:}
\end{tcolorbox}
\caption{Prompt used for evaluating Cultural Appropriateness level V3 in MLLM-as-Jury Evaluation}
\label{fig:prompt_cultural_v3}
\end{figure*}

\end{document}